\documentclass[aps,prd,twocolumn,groupedaddress,nofootinbib,amsmath,amssymb,floatfix]{revtex4-2}

\usepackage{graphicx}
\RequirePackage{hyperref}

\renewcommand{\vec}[1]{\mathbf{#1}}

\begin{document}
	
	\title{Perturbative quantum consistency near black-hole horizon formation}
	
	\author{Stefan Hofmann}
	\email{Stefan.Hofmann@physik.uni-muenchen.de}%Lines break automatically or can be forced with \\
	\affiliation{Arnold Sommerfeld Center for Theoretical Physics, Theresienstra{\ss}e 37, 80333 M\"unchen, Germany}
	
	\author{Maximilian Koegler}
	\email{M.Koegler@physik.uni-muenchen.de}%Lines break automatically or can be forced with \\
	\affiliation{Arnold Sommerfeld Center for Theoretical Physics, Theresienstra{\ss}e 37, 80333 M\"unchen, Germany}
	
	\author{Florian Niedermann}
	\email{Florian.Niedermann@su.se}%Lines break automatically or can be forced with \\
	\affiliation{Nordita, KTH Royal Institute of Technology and Stockholm University\\
		Hannes Alfv\'ens v\"ag 12, SE-106 91 Stockholm, Sweden}
	
	\date{\today}
	
	\begin{abstract}
		We study the prelude to black-hole formation using a suspended shell composed of physical matter that fulfills the dominant energy condition. Here, the collapse of the shell is brought to rest when the formation of the horizon is imminent but has not yet occurred. As the main achievement of this work, we obtain the Feynman propagator which connects the interior and the exterior of the shell within two local coordinate patches. It is derived by drawing an analogy to the propagation of light across interfaces that separate regions  with different susceptibilities inside a medium. As a first application, we use this propagator to determine the vacuum persistence amplitude in the presence of external sources. On timescales much shorter than the Page time, we find that the amplitude builds up with time yet remains consistent with perturbative unitarity. 
		
	\end{abstract}
	
	\maketitle
	
	\section{Introduction}
	\label{sec: intro}
	In recent years, observations have been able to test our understanding of black-holes with increasing accuracy. For example, the gravitational wave signal from a binary black-hole merger provides us with information about their masses and intrinsic vibrational modes~\cite{Abbott:2016blz}; similarly, the event horizon telescope~\cite{Akiyama:2019cqa} probes the classical geometry close to the event horizon. While all of these observations have been compatible with the prediction from general relativity, deep theoretical problems remain beyond the classical level. Most prominently, the information paradox arises from a semiclassical analysis of the formation and evaporation process of a black-hole. In a nutshell, it is based on the observation that an initially pure quantum state that collapses into a black-hole evolves into a mixed state after the black-hole has evaporated~\cite{Hawking:1976ra}, in contradiction to a consistent, unitary time evolution. 
	
	In this work, we study the quantum consistency of a thin-shell system close to black-hole formation. To this end, we will pursue a very conservative approach where we employ the standard semiclassical toolkit. This will avoid complications arising from the presence of a horizon, yet allow us to adiabatically probe the quantum stability of the shell system while it approaches a black-hole state. In other words, we start out with a configuration that is paradox-free and very slowly move toward horizon formation. We then ask if we can find any direct evidence for a buildup of quantum corrections that would indicate the breakdown of the semiclassical approximation. This could be hinting at a nonperturbative description of black-holes \cite{Page:1976df,  Giddings:2007ie, Akhmedov:2015xwa, Burgess:2018sou} as it has also been suggested through explicit constructions in the literature \cite{Lunin:2001jy,Mazur:2001fv, Lunin:2002qf,Dvali:2011aa, Dvali:2012wq,Almheiri:2012rt, Hofmann:2014jya,Gruending:2014rja,Bousso:2018bli, Almheiri:2019qdq,Penington:2019kki, Dvali:2020wft}.
	
	Our proposal faces two immediate challenges. First, we have to come up with a physical model that admits an arbitrarily slow  collapse. In particular, we cannot use standard collapse models where for typical black-hole masses the near-horizon regime as seen by a comoving observer is passed within microscopic timescales. Second, we have to find a sufficiently simple diagnostic tool, which enables us to probe the quantum evolution of fields on the shell background over long enough time intervals to be sensitive to accumulated growth effects.
	
	We overcome the first challenge by introducing a \textit{suspended shell} model. Here, the idea is to stabilize a massive, nonrelativistic shell with surface density $\rho$ through its surface pressure $p$, which we treat as a free parameter. Provided the shell radius $R$ is slightly larger than the Schwarzschild radius $r_g$, explicitly $R > 25 r_g /24$, we find that the shell can be brought to a complete stop while still being composed of matter fulfilling the strong and dominant energy condition. In such a static model, we can treat $R$ as a free dial to probe different stages of black-hole formation.  
	
	Regarding the second challenge, we will study the propagation of a quantum scalar field on the gravitational shell background. Specifically, we will calculate for an inertial observer the persistence amplitude of the Minkowski vacuum inside the shell in the presence of an external source~\cite{Schwinger:1953zza}. In particular, this diagnostic tool is sensitive to effects that buildup over time and could be missed in a standard stability perturbative analysis.\footnote{In~\cite{Akhmedov:2015xwa} and \cite{Burgess:2018sou}  a similar objective was pursued by studying secular growth close to Schwarzschild and Rindler horizons, respectively. Also, see \cite{Berczi:2020nqy} for a more recent work where coherent states are collapsed into black-holes. In contrast, our work has its focus on the Minkowski vacuum in the interior of the shell (rather than the asymptotic one) and our shell -- as seen by a comoving observer~-- is prevented from crossing the horizon. For a specific
microscopic realization of a suspended shell in the context of
string theory see \cite{Danielsson:2017riq}.} This calculation requires the Feynman propagator to be valid in the interior \textit{and} exterior to properly account for reflection and transmission effects at the shell. For simplicity, we will use a Riemann normal neighborhood (RNC) expansion anchored in the exterior vicinity of the shell. Together with the interior Minkowski spacetime, this flat patch provides a local covering of the physical manifold. This approximation enables us to derive an analytic result for the Feynman propagator and isolate its reflective and transmissive contributions. This is complementary to the gray body calculation as it is valid for high-frequency probes (as opposed to low frequency ones~\cite{Page:1976df}). It is also reminiscent of Hawking's analysis~\cite{Hawking:1974sw}, except that our collapsing body has been brought to a stop rather than being freely falling and our (local) vacuum will be defined inside the shell rather than asymptotically. 
	
	As a result of these investigations, we find that the vacuum persistence amplitude is enhanced by the presence of a near-critical shell. This effect gets stronger the longer the geometry is probed and the closer we get to horizon formation. Whether the amplitude would ever grow above unity, which would signal a quantum instability and hence a failure of the semiclassical treatment, cannot be answered conclusively within the validity of our approximation. The physical reason is that the exterior RNC patch has a finite temporal extent, which is hierarchically smaller than the lifetime of the black-hole as it corresponds to the inertial frame of a freely falling observer that reaches and crosses the shell in a short amount of proper time. However, we will learn how this challenge can be overcome in future work by using an orbital rather than a radially in-falling observer, who can then probe the shell indefinitely. 
	
	This article is organized as follows: In Sec.~\ref{sec:toyModel} we provide a warm-up by computing the propagation within two media with different susceptibilities. Later, the two media can be identified with the two coordinate patches inside and outside the shell. We present a perturbative technique, which describes one medium as an interaction term, as well as a nonperturbative approach based on matching conditions. Both techniques are shown to agree. In Sec.~\ref{sec:quantumCollapse}, we first introduce our suspended shell model and  then use it to compute the Feynman propagator, employing the same techniques as before.  Sec.~\ref{sec:QFTan} is then devoted to a calculation and discussion of the vacuum persistence amplitude.  We conclude in Sec.~\ref{sec:conclusion}. Throughout this article, we use the metric signature diag$(-,+,+,+)$ and units where $c=G=\hbar=1$.
	
	\section{Propagation Across Boundaries}
	\label{sec:toyModel}
	
	\subsection{Scalar field in a medium}
	
	Consider a massless and real scalar field $\phi$ coupled to an external source $J$ and placed in a spacetime-homogeneous medium in flat space. The effect of this medium on the field can be described in terms of a constant susceptibility $\varepsilon$, which influences the propagation and therefore the equation of motion
	\begin{align}
		\Box^\varepsilon \phi := \left( -\varepsilon \partial_t^2 + \partial_{\vec x}^2\right) \phi = -J\,.
		\label{eomScalarMed}
	\end{align}
	
	The corresponding dispersion relation for a plane wave $\exp\{\mathrm{i}(\omega^\varepsilon_kt - \vec{k} \vec{x})\}$ with momentum $\vec{k}$ is ${\omega_k^\varepsilon = \sqrt{\vec k^2}/\sqrt{\varepsilon}}$ in Cartesian coordinates. This effect can be fully captured by introducing for the scalar product an auxiliary metric ${\eta_\varepsilon = \mathrm{diag}(-1/\varepsilon,1,1,1)}$, for example used for the kinetic contractions on a Minkowski background $\mathcal M$ endowed with the metric~$\eta$. In the rest frame of the medium the action giving rise to \eqref{eomScalarMed} is composed of the free part~$S_0^\varepsilon$ and the interaction term $S_J$: 
	\begin{align}
		S_0^\varepsilon + S_J =  \int \limits_{\mathcal M} \mathrm d \mu(x) \left[ - \frac 1 2 \eta_\varepsilon^{\mu\nu} \partial_\mu \phi \partial_\nu \phi  +  J \phi   \right]  ,
		\label{kinMed}
	\end{align}
	with measure $\mathrm d \mu(x) := \mathrm d^4 x \sqrt{-\mathrm{det}(\eta)}$. In the limit $\varepsilon \rightarrow 1$ the usual Minkowski action is recovered.
	
	In this work, we are mainly interested in providing an explicit expression for the Feynman propagator, which is the time-ordered correlator evaluated in the vacuum state $| \Omega \rangle$,
	\begin{align}
		\Delta^{\varepsilon}_{xy}:=  \mathrm{i}\langle \mathrm T \phi_x \phi_y \rangle \, ,
		\label{greenDef}
	\end{align} 
	where T denotes time ordering, $ \langle \,.\, \rangle := \langle \Omega | \, . \, | \Omega \rangle$ and we use the shorthand ${f_x:=f(x)}$ and $f_{xy}:=f(x,y)$ for any function or distribution $f(x)$, $f(x,y)$, respectively. This propagator fulfills the fundamental equation of Green's functions
	\begin{align}
		\Box^\varepsilon \Delta_{xy}^\varepsilon = -\delta^{(4)}_{xy} \, ,
		\label{greenEqu}
	\end{align}
	with appropriate boundary conditions.
	
	Computing the propagator in such a medium with the correlator (\ref{greenDef}) then results in
	\begin{align}
		\Delta^\varepsilon_{xy} =  \theta {(x^t-y^t)} G^\varepsilon_{xy} +  \theta {(y^t-x^t)} G^\varepsilon_{yx}  ,
		\label{propEps}
	\end{align}
	where we introduced the Wightman distribution ${G^\varepsilon_{xy} :=\mathrm{i}\langle  \phi_x \phi_y \rangle}$ as
	\begin{align}
		G^\varepsilon_{xy} =  \int_k^\varepsilon \, {\rm e}^{\mathrm{i} k_\perp (x^\perp - y^\perp)} \, ,
		\label{particScalar}
	\end{align}
	with the shorthand
	\begin{align}
		\int_k^\varepsilon := \mathrm{i} \int \frac{{\rm d}^3 k }{(2\pi)^32\omega_k^\varepsilon \, \varepsilon} {\rm e}^{-\mathrm{i} \omega_k^\varepsilon (x^t -y^t)}{\rm e}^{\mathrm{i} \vec k_\parallel(\vec x^\parallel -\vec y^\parallel) } \, .
		\label{integralDef}
	\end{align}
	It is straightforward to check that $\Delta^\varepsilon$ fulfills Eq.~\eqref{greenEqu}. Here, the spacetime coordinates are labeled for later convenience as $x^\mu=\left( x^t,  x^\perp, \vec x^\parallel \right)$ and analogously for the spatial momenta $\vec k = (k_\perp, \vec k_\parallel)$. In the next sections we derive an expression for the propagator which is valid in the presence of regions with different susceptibilities.
	
	\subsection{Perturbative approach}
	\label{sub_toyPert}
	
	\subsubsection{Different dispersion relations encoded as interaction}
	
	We first present a perturbative derivation of the propagator in systems with boundaries. As an instructive example, we consider a spacetime region with domain $S$ and its complement $S^C$, characterized by the susceptibilities $\varepsilon_S$ and $\varepsilon$, respectively. We can think of such a system as a scattering object $S$ placed in a surrounding medium $S^C$. The free action of a scalar field in this system is given by
	\begin{align}
		S_0  =  - \frac 1 2  \int \limits_{S^C} \mathrm d \mu \,\, \eta_\varepsilon^{\mu\nu} \partial_\mu \phi \partial_\nu \phi  - \frac 1 2 \int \limits_{S} \mathrm d \mu \,\,  \eta_{\varepsilon_S}^{\mu\nu} \partial_\mu \phi \partial_\nu \phi  \, .
		\label{actionSO}
	\end{align}
	As before, the kinetic terms determine the dispersion relation of the scalar field within the scattering object and the surrounding medium. However, solving the defining equation of the propagator in such a system can become arbitrarily complicated depending on the geometry of the scattering object.
	
	This difficulty can be circumvented by adopting the following viewpoint. 
	We require that the field $\phi$ propagates everywhere according to the first kinetic term in (\ref{actionSO}); i.e., we identify it with the free action $S_0^\varepsilon$ introduced in (\ref{kinMed}). Since this leads to an error when describing propagation within the scattering object, we compensate for this by adding an interaction term $S_I$ to the action. This then corresponds to a formal (but fully equivalent) rewriting of (\ref{actionSO}); explicitly $S_0=S_0^\varepsilon+S_I$, where
	\begin{align}
		S_I := \int \limits_{S} \mathrm d \mu\, \mathcal{L}_I 
		:=- \frac 1 2\int \limits_{S} \mathrm d \mu  \left( \eta_{\varepsilon_S}^{\mu\nu} \partial_\mu \phi \partial_\nu \phi  - \eta_{\varepsilon}^{\mu\nu} \partial_\mu \phi \partial_\nu \phi \right).
		\label{actionNewVert}
	\end{align}
	Adopting this perspective, the propagator $\Delta^\varepsilon$ in (\ref{propEps}) describes the ``free'' propagation valid for a uniform medium with susceptibility $\varepsilon$. The effect of the scattering object is then encapsulated in the interaction term with Lagrangian $\mathcal L_{I}= \lambda_S  \left(\partial_t \phi\right)^2/2$ and coupling constant $\lambda_S = \varepsilon_S-\varepsilon$. Employing the interaction picture of field theory, the nonperturbative propagator in the presence of the scattering object, valid everywhere in space, is given by
	\begin{align}
		\!\Delta_{xy} = \mathrm{i}\left\langle \mathrm T \, \phi_x \phi_y \, \mathrm{ e}^{ \text{i}\frac{ \lambda_S}{2} \int\limits_{S} \mathrm d \mu_z   \left(\partial_{z^t} \phi_z\right)^2 } \right\rangle_{\!\!\!\mathrm{con}},
		\label{fullGreenF}
	\end{align}
	{where $\langle \cdot \rangle_{\rm con}$ only includes connected diagrams and as usual $\lim_{T_z \to \infty (1+\mathrm{i} \epsilon)} \int_{-T_z}^{T_z} \mathrm d z^t$ is understood.}
	Contractions of fields in this theory give rise to the propagator $\Delta^\varepsilon$ and $\varepsilon_S$ only occurs through the coupling constant~$\lambda_S$.
	
	We note that there is a formal symmetry under the exchange of $S$ and $S^C$. This implies that we could have instead defined the ``free'' propagation with respect to the region $S$ and the effect of the medium in $S^C$ in terms of an interaction term. In this way, $\Delta^{\varepsilon_S}$ would have been the propagator and $\varepsilon-\varepsilon_S$ the coupling constant. While this choice does not matter in the nonperturbative evaluation (\ref{fullGreenF}), it is relevant when (\ref{fullGreenF}) is truncated at finite order in $\lambda_S$.
	
	This is demonstrated by the following example: If we take the scattering object to cover the whole space, (\ref{fullGreenF}) gives rise to a geometric series that can be resummed to~$\Delta^{\varepsilon_S}$. This is not surprising, since in the same way a massive propagator can be generated from the resummation of the mass term $m^2 \phi^2$ ``interactions'' \cite{Schwartz:2013pla}. However, if one truncates at a finite order in $\lambda_S$, one does not obtain $\Delta^{\varepsilon_S}$, but a propagator which at best approximates~$\Delta^{\varepsilon_S}$. On the other hand, if one treats the spacetime region with susceptibility $\varepsilon$ as the interacting part, the integral in (\ref{fullGreenF}) has no support and thus vanishes. In this way, $\Delta^{\varepsilon_S}$ trivially emerges no matter how many orders in $\lambda_S$ we take into account. Therefore, for this particular system, the latter point of view is preferable. Moreover, if the difference between $\varepsilon$ and $\varepsilon_S$ is sufficiently small, evaluating (\ref{fullGreenF}) perturbatively to some order in $\lambda_S$ is a viable approach, but fails of course if the difference is too large. This nonperturbative regime necessarily sets in when $|\lambda_S|\geqslant1$ (or even earlier, as we will see below).
	
	\subsubsection{Reflection and transmission of a boundary}
	
\begin{figure}
		\includegraphics{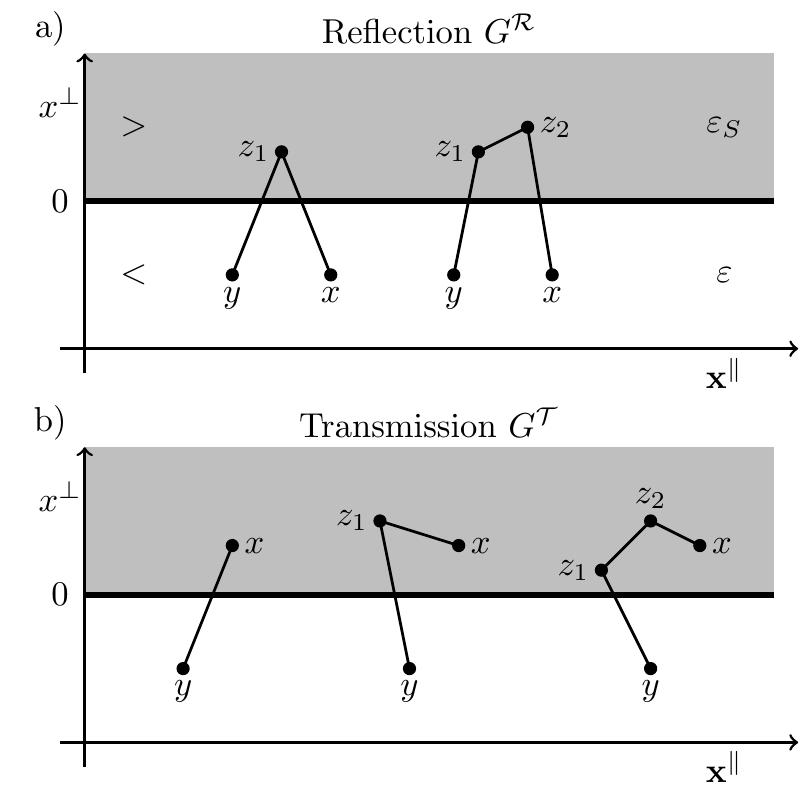}
		\caption{\label{figHalfSp} The scattering object with susceptibility $\varepsilon_S$ covering the upper half-space $>$ is shaded in gray, the lower half-space~$<$ with susceptibility $\varepsilon$ in white. The spatial projections of the contributions to the reflection propagator $G^{\mathcal{R}}$ and the transmission propagator $G^{\mathcal{T}}$ arising from (\ref{fullGreenF}) up to second order in $\lambda_S$ are represented schematically in a) and b), respectively.
		}
	\end{figure}

	Demonstrating the effects of a boundary separating two media with different susceptibilities can be best achieved for the simplest case of two half-spaces. The scattering object has susceptibility $\varepsilon_S$ and covers the upper half-space $>$ with spatial domain $S=\left\{ \vec x \,\, : \,\, x^\perp \geqslant 0 \right\}$. Correspondingly, the lower half-space $<$ with domain $S^C = \left\{ \vec x \,\, : \,\, x^\perp < 0 \right\}$ has susceptibility $\varepsilon$ as depicted in Fig.~\ref{figHalfSp}.
	
	We can expand (\ref{fullGreenF}) for small $\lambda_S$,
	\begin{multline}\label{perturbative_expansion}
		\Delta_{xy} =  \Delta_{xy}^{\varepsilon}  + \lambda_S  \int \limits_S \mathrm d \mu_{z_1} \, \partial_{z^t_1}\Delta^\varepsilon_{x z_1} \partial_{z^t_1} \Delta^\varepsilon_{z_1 y}  \\
		+ \, \lambda_S^2  \int \limits_S \mathrm d \mu_{z_1}  d \mu_{z_2} \, \partial_{z_1^t}\Delta^\varepsilon_{x z_1} \partial_{z_1^t}  \partial_{z_2^t}\Delta^\varepsilon_{z_1 z_2} \partial_{z_2^t} \Delta^\varepsilon_{z_1 y} + \mathcal{O}\left(\lambda_S^3 \right) .
	\end{multline}
	We then use standard field theory methods to further evaluate these different tree-level terms. This lengthy but straightforward calculation is detailed in Appendix~\ref{app2ndorder}. Fixing the causal arrangement with $x^t>y^t$ and setting $y^\perp<0$ for simplicity, the various contributions to the resummed propagator (or equivalently to the Wightman distribution) $G_{xy} := \Delta_{xy}|_{x^t>y^t}$ can be grouped in three terms,
	\begin{align}\label{G_decomposition}
		G_{xy} =  \left( 1 - \theta_{x^\perp} \right) \left( G^\varepsilon_{xy} + G^{\mathcal{R}}_{xy} \right) + \theta_{x^\perp}  G^{\mathcal{T}}_{xy} \, .
	\end{align}
	There is a single term $G^\varepsilon_{xy}$, as defined in \eqref{particScalar}, describing direct propagation from $y$ to $x$ in the lower half-space $<$. The two diagrams depicted in Fig.~\ref{figHalfSp}a) contribute to the reflection propagator (for $x^\perp < 0$)
	\begin{align}
		G^{\mathcal{R}}_{xy} =  -\int_k^\varepsilon \, {\rm e}^{\mathrm{i} k_\perp (x^\perp + y^\perp)} \mathcal{R}(\bar \omega) \, ,
		\label{reflExpand}
	\end{align}
	with reflection coefficient
	\begin{align}
		\mathcal{R}(\bar \omega) &=  \frac{1}{4} \bar \omega - \frac{1}{8} \bar \omega^2 + \mathcal O\left( \bar \omega^3 \right)  \, ,
		\label{reflexExpand}
	\end{align}
	where for convenience we introduced a rescaled expansion parameter $\bar \omega=\lambda_S{ (\omega_k^\varepsilon)^2}/{k_\perp^2} ={\lambda_S}[1+\tan^2(\alpha)]/{\varepsilon}$ with the angle of incidence $\alpha$ defined through $\vec k_\parallel^2 =:~ \tan^2 (\alpha) k_\perp^2$. The remaining terms, as shown in Fig.~\ref{figHalfSp}b), give rise to the transmission propagator (for $x^\perp \geqslant 0$)
	\begin{multline}
		G^{\mathcal{T}}_{xy}  =\int_k^\varepsilon \, {\rm e}^{\mathrm{i} k_\perp (x^\perp - y^\perp)}  \mathcal{T}(\bar \omega)  \\
		\times \Bigg[ 1 +\frac{\mathrm{i}}{2} \left(\bar \omega - \frac{\bar \omega^2}{4}\right) k_\perp x^\perp
		- \frac{1}{8} \bar \omega^2 k_\perp^2 \left(x^\perp\right)^2  \Bigg] + \mathcal{O} \left(\bar \omega^3 \right) \,, \label{transmission_prop_2n_order}
	\end{multline}
	which we have factorized into $x^\perp$ dependent and independent terms. The latter are collected in the transmission coefficient
	\begin{align}
		\mathcal{T}(\bar \omega) = 1 -\frac{1}{4} \bar \omega + \frac{1}{8} \bar \omega^2 + \mathcal O\left( \bar \omega^3 \right) = 1 - \mathcal{R}(\bar{\omega}) \, .
		\label{transExpand}
	\end{align}
	The $x^\perp$ dependent terms, on the other hand, can be resummed as an exponential, giving rise to  the final expression (for $x^\perp \geqslant 0$)
	\begin{align}
		G^{\mathcal{T}}_{xy}  = \int_k^\varepsilon \, {\rm e}^{\mathrm{i} q_\perp x^\perp - \mathrm{i} k_\perp y^\perp} \mathcal{T}(\bar \omega) \, ,
		\label{transFull}
	\end{align}
	where
	\begin{align}
		\quad q_\perp(\vec k) = {\rm sgn}(k_\perp) \sqrt{ \left( \frac{\varepsilon_S}{\varepsilon} - 1 \right) \vec k_\parallel^2 + \frac{\varepsilon_S}{\varepsilon} k_\perp^2} \, .
		\label{momRelScalar}
	\end{align} 
	The replacement $k_\perp \to \, q_\perp$ ensures that the dispersion relation inherited from $S$ holds in the upper half-plane, i.e. $\Box_x^{\varepsilon_S} G^{\mathcal{T}}_{xy} =0$ (whereas $\Box_x^{\varepsilon} G^{\mathcal{R}}_{xy} =0$ in the lower half-plane).
	
	If we want to truncate the $\bar{\omega}$ expansion and work with a finite number of terms, we must guarantee the smallness of $\bar \omega$ by requiring $\lambda_S/\varepsilon \ll 1$ and $\tan^2(\alpha) \ll 1$. Therefore, in the perturbative regime, only scenarios with $\mathcal{T} \approx 1$ and $\mathcal{R} \ll1 $ can be consistently described. Since we are later interested in systems where total reflection can occur, a closed-form expression of this series would be desirable. Fortunately, for the case at hand a resummation is feasible and given by
	\begin{align}
		\mathcal{T}(\bar \omega) &= \sum_{n=0}^\infty \frac{ \sqrt \pi }{(1+n)! \Gamma(\frac{1}{2}-n)} \bar \omega^n = \frac{2}{1+\sqrt{1 + \bar \omega}}  \nonumber\\
		& = 1 - \mathcal{R}(\bar{\omega}) \, ,
		\label{transCoeff}
	\end{align}
	where $\Gamma$ denotes the Gamma function. {For the first three terms in the sum, the Gamma function becomes $\sqrt{\pi}, -2 \sqrt{\pi}$ and $4 \sqrt{\pi}/3$ giving the transmission coefficient as \eqref{transExpand}.}\footnote{We have checked this result only up to second order in $\bar{\omega}$ but a complementary nonlinear derivation will consolidate it further.}
	The closed-form expression in terms of the angle of incidence $\alpha$ is computed to be
	\begin{align}
		\mathcal{R}(\alpha)    =  \frac{\sqrt{ \left({\frac{\varepsilon_S}{\varepsilon}}-1 \right) \tan^2(\alpha)+{\frac{\varepsilon_S}{\varepsilon}}}-1}{\sqrt{ \left({\frac{\varepsilon_S}{\varepsilon}}-1 \right) \tan^2(\alpha)+{\frac{\varepsilon_S}{\varepsilon}}}+1} \, .
		\label{refCoeff}
	\end{align}
	
	A few remarks are in order. First, the Feynman propagators describing reflection and transmission can be obtained as usual by time ordering, e.g.  ${\Delta^{\mathcal{R}|\mathcal{T}}_{xy} = \, \theta(x^t-y^t) G^{\mathcal{R}|\mathcal{T}}_{xy} + \theta(y^t-x^t) G^{\mathcal{R}|\mathcal{T}}_{yx}}$. Second, identifying the series through the first two orders could only be achieved due to the simplicity of the boundary. For example for systems with a geometrically more complicated boundary one has to stick to the perturbative expansion in (\ref{perturbative_expansion}).  Third, the resummed reflection and transmission coefficients fulfill the identity $\mathcal{R}(\bar \omega) + \mathcal{T}(\bar \omega) = 1$ and agree with the spin-averaged Fresnel equations. They are also correct if $\lambda_S/{\varepsilon}$ is large. In fact, total reflection, $ |\mathcal{R}|^2 \rightarrow 1$, does occur for ${\varepsilon_S}/{\varepsilon} \rightarrow \infty$ (or $\lambda_S/\varepsilon \rightarrow \infty$ equivalently), describing a perfect mirror. Alternatively, for $\varepsilon>\varepsilon_S$ total reflection can also be achieved in the two cases ${\varepsilon_S}/{\varepsilon} \rightarrow 0$ and $\alpha \rightarrow \pm {\pi}/{2}$. Both scenarios correspond to total internal reflection, which in optical experiments is only caused by the latter limit though, i.e.~for a large enough angle of incidence $\alpha$. With all sources of total reflection in this optical example revealed, we can later address these as an analogy in the context of gravitational collapse.

	In summary, spacetime regions with different dispersion relations can be incorporated on the level of the action by introducing a bilinear interaction term with limited spacetime support. A propagator describing propagation across boundaries is then derived by using the interaction picture of quantum field theory.

	\subsubsection{Interlude: Double-slit Experiment}
	
	This procedure is not restricted to the basic setup presented here but can be extended to say diffraction experiments. As an instructive example, the propagator for the prominent double-slit experiment is obtained by taking an infinitely thin and opaque scattering object with domain $S=\left\{ \vec x \,\, : \,\, f(\vec x^\parallel) = 0 \wedge  x^\perp = 0 \right\}$, where $f(\vec x^\parallel)$ is the aperture, which is 1 on the intervals of the slits and 0 elsewhere. The susceptibility of $\varepsilon_S$ is assumed to be large to ensure that there is no propagation through and within the scattering object, which is a typical assumption for diffraction experiments. As before, the transmission propagator is obtained with (\ref{fullGreenF}). 
	
	The support of the integral in (\ref{fullGreenF}) can be conveniently implemented through $\delta({z^\perp}) [1-f(\vec x^\parallel)]$. Written this way, the first term describes an opaque scattering object without holes and thus the transmission vanishes. The second term, however, contributes to transmission but constrains the intermediate points $\vec z$ in (\ref{fullGreenF}) to lie within the aperture $F=\left\{ \vec z \,\, : \,\, f(\vec z^\parallel) = 1 \wedge  z^\perp = 0 \right\}$. Thus, the dynamics are dictated by the propagation of the fields from the lower half-space through the aperture to the upper half-space in accordance with Fraunhofer diffraction. The transmission propagator then becomes (for $x^\perp \geqslant 0$)
	\begin{align}
		\Delta^{\mathcal{T}}_{xy} = \varepsilon \int \limits_F \mathrm d \mu_{z_1} \, \partial_{z_1^t}\Delta^\varepsilon_{x z_1} \partial_{z_1^t} \Delta^\varepsilon_{z_1 y}  \;.
	\end{align}
	As before, the exact treatment with resummation is possible due to simplifying approximations. In the next section, we follow an approach that does not rely on resummation and can still provide a nonperturbative result.
	
	\subsection{Nonperturbative approach}\label{sec:toy_np}
	
	\subsubsection{Matching conditions across boundaries}
	
	For the case of a boundary separating two half-spaces with different susceptibilities, as depicted in Fig.~\ref{figHalfSp}, an exact solution of the tree-level Feynman propagator $\Delta_{xy}$ was found by expanding and resumming the interaction exponential in the correlator (\ref{fullGreenF}). Here, we present a different approach which does not rely on a perturbative expansion. To this end, we use Eq.~\eqref{transExpand} and Eq.~\eqref{reflExpand} as an ansatz with a priori arbitrary coefficients $\mathcal{T}(k)$ and~$\mathcal{R}(k)$, which are then fixed by matching conditions across the boundary. To be specific, continuity of $\phi$ and its normal derivative $n^{\mu}\partial_\mu \phi$ implies that $\Delta_{xy}$ and  $n^\mu \partial_\mu \Delta_{xy}$ are continuous across the boundary. Here, we introduced the boundary normal vector $n = \partial_{x^\perp}$.
	
	With the source in the lower half-space $<$, i.e.~ for $y^\perp < 0$, there are two contributions to $\phi|_{x^\perp<0}$: one corresponding to direct propagation between source and detector described in terms of $\Delta^\varepsilon$, and one due to the reflection off the boundary described in terms of $\Delta^\mathcal{R}$. Combining both, the full propagator defined exclusively in region $<$ becomes $\Delta^< := \Delta^\varepsilon+\Delta^\mathcal{R}$. Using (\ref{particScalar}) and (\ref{reflExpand}), we find (assuming again $x^t > y^t$)
	\begin{align}
		G^<_{xy} =  \int_k^\varepsilon \left[{\rm e}^{\mathrm{i} k_\perp (x^\perp - y^\perp)}  -  {\rm e}^{-\mathrm{i} k_\perp (x^\perp + y^\perp)} \mathcal{R}(\vec k)\right]  .
		\label{reflectionScalar}
	\end{align} 
	Correspondingly, in region $>$ there is only one contribution for the resulting field configuration due to the transmission part (\ref{transFull}). Thus, the propagator for region $>$ reads
	\begin{align}
		G^{>}_{xy} = \int_k^\varepsilon  {\rm e}^{\mathrm{i} q_\perp x^\perp-\mathrm{i} k_\perp y^\perp} \,  \mathcal{T}(\vec k) \, ,
		\label{transScalar}
	\end{align}
	where $q_\perp(\vec{k}_\parallel, k_\perp)$ is given in (\ref{momRelScalar}).
	
	With the propagators of the two half-spaces the boundary conditions explicitly translate to
	\begin{subequations}
		\label{bndryCond}
		\begin{align}
			\lim_{x^{\perp} \nearrow \, 0} G^{<}_{xy}&= \lim_{x^{\perp} \searrow \, 0} G^{>}_{xy} \, , \label{bndryCond1} \\
			\lim_{x^{\perp} \nearrow \, 0}\partial_{x^\perp} G^{<}_{xy} &= \lim_{x^{\perp} \searrow \, 0} \partial_{x^\perp} G^{>}_{xy} \label{bndryCond2} \,. 
		\end{align}
	\end{subequations}
	Inserting the Green functions (\ref{reflectionScalar}) and (\ref{transScalar}), we obtain
	\begin{align}
		1 - \mathcal{R}(\vec k) &= \mathcal{T}(\vec k) \, , \qquad k_\perp + k_\perp \mathcal{R}(\vec k) = q_\perp \mathcal{T}(\vec k) \,.
	\end{align}
	Solving for the reflection and transmission coefficient results in
	\begin{align}
		\mathcal{R}(\vec k)&=  \frac{q_\perp-k_\perp}{k_\perp+q_\perp} \, , \qquad  \mathcal{T}(\vec k)= \frac{2k_\perp}{k_\perp+q_\perp} \, ,
		\label{tHalfM}
	\end{align}
	which can be shown to agree with the perturbatively found coefficients (\ref{transCoeff}) and (\ref{refCoeff}) by using the momentum relation (\ref{momRelScalar}). Taking Eq.~\eqref{transExpand} and Eq.~\eqref{reflExpand} as an ansatz together with the boundary conditions (\ref{bndryCond}) therefore constitutes an alternative for deriving exact propagators across boundaries. Since the feasibility of this approach heavily depends on the shape and motion of the boundary, the perturbative method can always serve as a fallback for systems more complicated than the two half-space system.

	\subsubsection{Conservation of charge density current}
	\label{consEsec}
	To further consolidate our formalism, we study the charge flux across the boundary. For this purpose, we compare the current of a complex scalar field $j_\mu[\phi]=\mathrm{i} \left( \phi^* \partial_\mu \phi - \partial_\mu \phi^* \phi \right)$ on both sides of the boundary and use the continuity of the normal derivative of $j$ across the boundary. For the system of two half-spaces we discussed above, this amounts to $\eta^{\mu\nu}_\varepsilon n_\mu j_\nu[\phi^<] \big|_{z^\perp \nearrow \, 0} =~\eta^{\mu\nu}_{\varepsilon_S} n_\mu j_\nu[\phi^>]\big|_{z^\perp \searrow \, 0}$. As an example, we couple the scalar field to the external source $J(k) \propto \delta^{(3)}(\vec k-\vec {\tilde k})$, generating a monochromatic plane wave with spatial momentum $\tilde{\vec k}$, and compute the resulting scalar field in the respective half-space with ${\phi^{<|>}_x=\int \mathrm d \mu_y \, \Delta^{<|>}_{xy} J_y}$. Normalizing the initial current to~$1$, the reflected and transmitted part lead to the following equality:
	\begin{align}
		1 = \left| \mathcal{R}(\tilde k_\perp) \right|^2 +  \frac{\mathrm{Re}\left[q_\perp(\tilde k_\perp)\right]}{\tilde k_\perp} \left| \mathcal{T}(\tilde k_\perp) \right|^2 \, .
		\label{chargeDensCons}
	\end{align}
	Inserting the reflection and transmission coefficient in~\eqref{tHalfM}, we see this relation is indeed fulfilled. This shows that the propagators in (\ref{reflectionScalar}) and (\ref{transScalar}) are compatible with a conserved charge flux across the boundary. Physically, the first term on the rhs., $|\mathcal{R}|^2$, corresponds to the current's fraction that is being reflected, and the second term, $\mathrm{Re}(q_\perp) |\mathcal{T}|^2/k_\perp$, to the transmitted part. We will refer to these observables as the ``reflectance'' and ``transmittance'', respectively.  
	
	Having established a perturbative and a nonperturbative technique to calculate propagators across boundaries, we can apply these to a gravitational system with boundaries in the next section.
	
	\section{Propagation across shells}
	\label{sec:quantumCollapse}
	
	\subsection{Gravitational setting}
	
	\subsubsection{Schwarzschild geometry}
	
	Following Birkhoff's theorem, any compact static gravitational source of mass $M$ results in a Schwarzschild spacetime for an observer sufficiently far away \cite{ssbirkhoff}. The line element of this geometry in spherical Schwarzschild coordinates $(t_S,r,\vartheta,\varphi)$ is given by
	\begin{align}
		\mathrm d s^2 = - f(r)\, \mathrm d t^2_S + f^{-1}(r) \,\mathrm d r^2 + r^2 \mathrm d \Omega_2 \, ,
		\label{SchSLine}
	\end{align}
	where $f(r) =  1 - {r_g}/{r} $ with the Schwarzschild radius $r_g=2M$ and $\mathrm d \Omega_2 =  \mathrm d \vartheta^2 + \sin^2(\vartheta) \,\mathrm d \varphi^2 $. By construction the exterior region $r>r_g$ is static relative to the corresponding observer field $u_S= f^{-1/2} \partial_{t_S}$ (suppressing the $r$ dependence of $f$). The trajectories of the Schwarzschild observer are the integral curves $\gamma$ of $u_S$. The geodesic equation yields $\ddot \gamma = r_g/2r^2 \, \partial_r$, which is the acceleration an observer must sustain to stay at rest. 
	
	It is often convenient to anchor coordinates to a specific family of freely falling observers. Since (\ref{SchSLine}) exhibits a coordinate singularity at $r=r_g$, we choose an observer for which the line element remains finite at $r=r_g$. Specifically, we consider a freely falling observer with eigentime $\tau$ and {velocity $u$} that is at rest for $r\rightarrow\infty$. Introducing its radial position $R(\tau)$ and demanding $g(u,u)=-1$, we derive its velocity as {$u(R)=f^{-1}(R) \partial_{t_S} -\sqrt{1-f(R)} \partial_r$}, where the sign in the radial component has been chosen in accordance with an in-falling observer. The dual vector field can be written as $u^\star(R) = - \text{grad}[t_P(t_S,r)]|_{r=R}$ with $t_P(t_S,r) = t_S + \int^r_0 \mathrm d r' u^\star_r(r')$. Rewriting the line element (\ref{SchSLine}) by using $\mathrm d t_S = \mathrm d t_P - u^\star_r(r) \mathrm d r$ yields the Schwarzschild geometry in Painlev\'e-Gullstrand coordinates $(t_P,r,\vartheta,\varphi)$,
	\begin{align}
		\! \mathrm d s^2 = - f\, \mathrm d t^2_P + 2 \sqrt{1-f} \mathrm d t_P \mathrm d r + \mathrm d r^2 + r^2 \mathrm d \Omega_2 \, .
		\label{PGLine}
	\end{align}
	In these coordinates the Schwarzschild geometry is manifestly regular at $r=r_g$ and the velocity of the freely falling observer is given by {$u(R) = \partial_{t_P} -\sqrt{1-f(R)} \partial_r$}.

	\begin{figure}
		\includegraphics{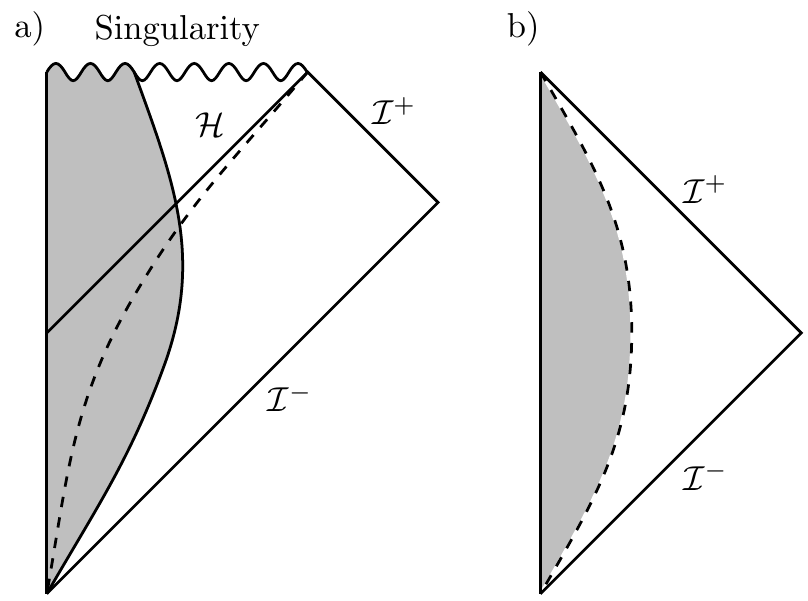}
		\caption{\label{penroseCollapse} Schematic Penrose diagram for the collapsing shell in a) with the Minkowski metric inside the shell depicted as the gray area. The exterior geometry is Schwarzschild with the horizon $\mathcal H$ at $r=r_g$ and singularity at the end of the collapse. Additionally, one constant radius $R>r_g$ is indicated by a dashed line. In b) the Penrose diagram for a fixed shell of said radius $R$ is sketched again with the Minkowski metric inside (gray area) and Schwarzschild outside (white area) which does not possess a horizon.}
	\end{figure}

	\subsubsection{Shell geometry}
	\label{subsub_shellGeo}

	Historically, black-hole formation was first studied for a collapsing dust cloud, where dust refers to the lack of interactions between the particles within the cloud except of gravity \cite{PhysRev.56.455}.
	Since most applications do not depend on the details of the collapsing star model such as in \cite{PhysRevLett.14.57, hawking1975particle}, a convenient alternative is to consider the system of a collapsing shell. This system possesses a Minkowski spacetime inside and Schwarzschild geometry outside as depicted in Fig.~\ref{penroseCollapse}a). To be precise, the full line element is
	\begin{align}\label{shell_geom}
		\mathrm{d}s^2 = 
		\begin{cases}
			- \mathrm (d x^t)^2 +  \mathrm d r^2 + r^2 \mathrm d \Omega_2 =:d s^2_< , & r < R\\
			- f(r)\, \mathrm d t^2_S + f^{-1}(r) \,\mathrm d r^2 + r^2 \mathrm d \Omega_2=:d s^2_>.\!\! &  r > R
		\end{cases}
	\end{align} 
	The time coordinate is not continuous across the shell, i.e., the Minkowski time coordinate $x^t$ in the interior of the shell is different from $t_P$ and $t_S$ in (\ref{SchSLine}) and (\ref{PGLine}). We derive their relation in Appendix~\ref{subsec: appendix Israel} by matching the exterior line element $\mathrm ds^2_>$ and the interior line element $\mathrm d s^2_<$ across the shell and find in accordance with \cite{Israel:1966rt} 
	\begin{align}
		\frac{\mathrm d t_S}{\mathrm (d x^t)^2} = \frac{\sqrt{f(R) + \dot R^2}}{f(R) \sqrt{1 + \dot R^2 }} \, ,
		\label{insideTimeOutside}
	\end{align}
	where $\dot R = \mathrm d R/ \mathrm d \tau$. Here, $\tau$ is the proper time of the shell with intrinsic geometry
	\begin{align}
		\mathrm{d} s^2_{\text{Shell}} = -\mathrm{d} \tau^2 + R(\tau)^2 \mathrm {d} \Omega_2 \, .
	\end{align}

	A junction condition follows from Einstein's field equations across the shell as discussed in Appendix~\ref{subsec: appendix Israel}. It relates the discontinuity across the shell of the extrinsic curvature tensor to the shell-localized energy-momentum tensor,  $S_{\mu\nu}$. Here, we model $S_{\mu\nu}$ as a perfect fluid with energy density $\rho$ and surface pressure $p$, explicitly $S^\mu_\nu= \text{diag}(-\rho,p,p)$. This leads to the following set of equations, 
	\begin{align}
		\rho = \frac{\sqrt{1+ \dot R^2} - \sqrt{f(R)+ \dot R^2}}{4 \pi R}, \quad p = - \frac{R}{2 \dot R} \dot \rho - \rho  \, .
		\label{shellPre} 
	\end{align}
	If we commit to a particular matter model by fixing the equation of state parameter, the above equation will determine all three functions $\rho(\tau)$, $p(\tau)$ and the shell trajectory $R(\tau)$ up to initial conditions. In this work, however, we follow a different path. We first demand a particular trajectory $R(\tau)$ and then check whether it can be realized in terms of physical matter that fulfills the standard energy conditions. The simplest example, which will also be our main working model, corresponds to a shell at rest -- or a suspended shell -- with $\dot R = 0$. {For $R>r_g$ this spacetime does not possess a horizon as can be seen in the Penrose diagram in Fig.~\ref{penroseCollapse}b).} In this case, we have
	\begin{align}
		\rho = \frac{1-\sqrt{f(R)}}{4 \pi R}, \quad p= \frac{2R\left(1- \sqrt{f(R)}\right)-r_g}{16 \pi \sqrt{f(R)}R^2} \, .
		\label{fixShellRhoPre}
	\end{align}
	While for sufficiently large radii $R$ all energy conditions are fulfilled; for $R \leq 25 r_g/24$, on the other hand, the dominant energy condition $\rho \geq |p|$ is violated. Since this condition guarantees subluminal flow within the perfect fluid, such a shell cannot be stabilized in terms of a standard matter model (at least, this would require some more exotic microscopic model giving rise to an equation of state parameter greater than $1$). In the limit $R \to r_g$, the pressure even diverges, which tells us that an infinite force would be needed to hold the shell in place. Therefore, a fixed shell is a fully consistent geometry only for $R > 25 r_g/24$. For $R \leq 25 r_g/24$, on the other hand, nonstandard matter is needed, which, at the latest, becomes unphysical very close to horizon formation where $p \rightarrow \infty$. A different question concerns the stability of the setup under general metric perturbations, which might further tighten the constraint on $R$.
	
	For completeness, we also consider the scenario of a shell that follows the trajectory of the freely falling Painlev\'e-Gullstrand observer with $\dot R = - \sqrt{1-f(R)}$. Inserting this velocity in \eqref{shellPre}, we find
	\begin{align}
		\rho &= \frac{\sqrt{f_+(R)}-1}{4\pi R }, \quad  p = \frac{1}{8 \pi R} \left(1 + \frac{r_g- 2R f_+(R)}{ 2R \sqrt{f_+(R)}} \right) ,
		\label{PGShellRhoPre}
	\end{align}
	with $f_+({R})=1+r_g/{R}$.
	In contrast to our suspended shell model, both quantities are regular and finite before and during horizon crossing at $R=r_g$. Furthermore, the strong and dominant energy conditions are fulfilled throughout the entire collapse, which makes this model universally applicable. The pressure is always negative, which compared to a dust shell with $p=0$ leads to a quicker collapse. This should be contrasted with the suspended shell for which a positive pressure was needed to stabilize it. In any event, in the following, we will determine the Feynman propagator exclusively for the case where the shell is at rest, i.e.\ $\dot R =0$. 
	
	\subsection{Perturbative approach}
	\label{sec:curvePert}

	Our starting point is the theory of a scalar field in the suspended shell background \eqref{shell_geom}
	\begin{multline}
		S =  - \frac 1 2\int_< \mathrm d^4 x \sqrt{-\det(\eta)} \,\, \eta^{\mu\nu} \partial_\mu \phi \partial_\nu \phi
		\\- \frac 1 2\int_> \mathrm d^4x \sqrt{-\det(g_>)} \,\,   g_>^{\mu\nu} \partial_\mu \phi \partial_\nu \phi  \, .
	\end{multline}
	We will see that the effect of a curved background on the propagation of a scalar field, like for the optical case, can be captured in terms of a spacetime-dependent interaction term. To that end, we define the undisturbed propagation with respect to the Minkowski spacetime inside the resting shell. This choice also defines the vacuum of our local scattering experiment. The corresponding free action is that of a massless scalar field on a Minkowski background integrated over \textit{all} of space,
	\begin{align}
		S_0 =  - \frac 1 2\int \mathrm d \mu(x) \,\,  \eta^{\mu\nu} \partial_\mu \phi \partial_\nu \phi \, ,
		\label{actionMinkIns}
	\end{align}
	with Minkowski measure $\mathrm d \mu(x) = \mathrm{d}^4x \sqrt{-\det(\eta)}$. To construct the interaction term, we express the Schwarzschild geometry (\ref{SchSLine}) in terms of the time inside the shell $x^t(t_S)$ through  \eqref{insideTimeOutside} and set $\dot R = 0$. This yields
	\begin{align}
		\mathrm d s_>^2 
		&=- \frac{f(r)}{f(R)}\, \mathrm (d x^t)^2 + f^{-1}(r) \,\mathrm d r^2 + r^2  \mathrm{d} \Omega_2\, ,
		\label{shellOutMetric}
	\end{align}
	where the subscript $>$ indicates that $r \geq R$.
	
	Like in Sec.~\ref{sub_toyPert}, we treat the difference between the action in the exterior and the ``Minkowski action'' in (\ref{actionMinkIns}) as an interaction term with support only in the exterior. The full action becomes $S=S_0 + S_\mathrm{int}$, with
	\begin{align}
		S_\mathrm{int} := \frac{1}{2} \int_> \mathrm d \mu \, g_I^{\mu\nu} \partial_\mu \phi \partial_\nu \phi \, ,
	\end{align}
	where we introduced the auxiliary interaction metric, which is only defined in the exterior and reads
	\begin{align}
		\!\!\!	g_I^{\mu\nu} :=  \sqrt{\frac{\det(g_>)}{\det(\eta)} }g_>^{\mu\nu}  - \eta^{\mu\nu}   \, .
		\label{intPG}
	\end{align}
	Note that, as in the optical scenario, it is not a genuine metric, because it does not satisfy the Einstein field equations nor does it transform like a second rank tensor. To reiterate our previous point, this is merely a trivial rewriting of the action of a free scalar field living on the background of a shell at rest. For the perturbative expansion to work, we introduce
	\begin{align}\label{lambda}
		\lambda =\frac{r_g}{R} \ll 1 
	\end{align}
	as our smallness parameter. This approximation prevents us from examining a shell radius that is too close to $r_g$. This will be relaxed later when we work nonperturbatively in $\lambda$ in Sec.~\ref{sec:non_pert}. We further assume that the propagator will be evaluated close to the shell within a cubic box of edge length $\ell_0 \ll R$. This suggests introducing Cartesian coordinates anchored at the shell,
	\begin{align}\label{x_coord}
		(x^\perp, \vec{x}^\parallel) =  ( r - R, \, r \vartheta \cos(\varphi), \, r \vartheta \sin(\varphi))\,,
	\end{align}
	where $(|x^\perp|, |\vec{x}^\parallel|) < (\ell_0/2,\ell_0/2)$, and we used that \mbox{$\vartheta < \ell_0/(2R )\ll 1$}. We can then use $\bar{x}^\perp = x^\perp/R $ as a second smallness parameter. To be precise, we will employ a double expansion scheme where $\lambda \ll \bar x^\perp  \ll 1$.  Going in \eqref{intPG} up to order $\lambda \bar x^\perp$, we find
	\begin{multline}\label{g_I_2}
		\!\!\!  g_I(x) \simeq \frac{\lambda}{2} \mathrm{diag}\Big[ - \left(1 - 2 {\bar x^\perp} + \frac{3 \lambda}{4} + 2 (\bar x^\perp)^2 - 3 \lambda \bar x^\perp \right),\\
		-1 - 2 {\bar x^\perp} - \frac{\lambda}{4} - 2 (\bar x^\perp)^2 + \lambda \bar x^\perp, 1+\frac{3\lambda}{4} , 1+\frac{3\lambda}{4} \Big],
	\end{multline}
	which is valid up to corrections of order $\lambda^2$. Comparing this with the interaction metric in  \eqref{actionNewVert} shows that there is no perfect one-to-one correspondence between the optical setup and the suspended shell case. In particular, all diagonal elements are now nonvanishing and both the $tt$ and $\perp \perp$ components picked up an explicit spatial dependence. 
	
	The Feynman propagator can be calculated as we did in Sec.~\ref{sub_toyPert} by expanding the closed-form expression
	\begin{align}
		\!\Delta_{xy} = \mathrm{i}\left\langle \mathrm T \phi_x \phi_y \, \mathrm{ e}^{ - \frac{\text{i}}{2} \int\limits_{S} \mathrm d \mu_z  (g_I^{\mu\nu})_z {\partial_{z^\mu}} \phi_z {\partial_{z^\nu}} \phi_z + B } \right\rangle_{\!\!\!\mathrm{con}},
		\label{fullGreenFpg}
	\end{align}
	where we added the boundary term
	\begin{align}
		B = -\frac{\mathrm i}{2} \int \mathrm{d}\tilde \mu_z \left( \phi \, n^\mu_<\nabla_\mu \phi-\phi \, n^\mu_>\nabla_\mu \phi  \right)\,,
	\end{align}
	with $\mathrm{d}\tilde \mu_z$ denoting the (Minkowski) shell surface element and the outward-pointing shell normal vectors obtained from (\ref{normVec}) for the suspended shell, i.e. $\dot R = 0$,
	\begin{subequations}
		\label{normVec_suspended}
		\begin{align}
			n_>^\mu(R) &=\left(0, \sqrt{g_>^{\perp \perp}}, 0, 0\right) \, ,\\
			n_<^\mu(R) &=\left(0, 1, 0, 0\right)\,.
		\end{align}
	\end{subequations}
	As opposed to the optical investigation, $B$ contributes in general to the amplitude because the interacting term probes the direction perpendicular to the shell denoted by $\partial_{z^\perp}$.

	At linear order in $\lambda$ (corresponding to linear order in $g_I$), we obtain
	\begin{align}
		&\Delta_{xy} =  \Delta_{xy}^{\eta}  -  \int  \mathrm d \mu_{z} \, (g_I^{\mu\nu})_z \partial_{z^\mu}\Delta^{\eta}_{x z} \partial_{z^\nu} \Delta^{\eta}_{z y} \nonumber \\
		&-\frac{1}{2}  \int \mathrm{d}\tilde \mu_z \!\! \left[ \sqrt{\frac{(g_>^{\perp \perp})_z}{f(R)}}-1\right] \!\! \left[ \Delta_{xz}^{\eta} \partial_{z^\perp}  \Delta_{zy}^{\eta} + \Delta_{zy}^{\eta} \partial_{z^\perp}  \Delta_{xz}^{\eta}\right] ,  \label{perturbative_expansionRNC} 
	\end{align}
	where $\Delta^{\eta}_{x y}$ is the Feynman propagator in Minkowski space. 
	There is one new subtlety involved. The intermediate integration $\int \mathrm d \mu_z$ still extends to infinity which seems to invalidate our approximation that ${|(z^\perp, \vec{z}^\parallel)|\ll R}$. This, however, is not a problem provided $|x^t-y^t|\ll \ell_0$ and $|(x^\perp - y^\perp, \vec{x}^\parallel - \vec{y}^\parallel)|\ll \ell_0$ as the causal structure of the theory then implies that $\int \mathrm{d}\mu_y \Delta_{xy} J_y$ is not probing the space outside the box of size $\ell_0$. In other words, contributions to \eqref{perturbative_expansionRNC} where $  |( z^\perp,\vec{z}^\parallel)| > \ell_0$ do not contribute.\footnote{Strictly speaking, Feynman propagation is not localized on the light cone and hence probes spacetime arbitrarily far away. These contributions are however strongly suppressed for the separation of scales considered here. Alternatively, we could have introduced a cutoff for the configuration space integration and demonstrated that observables do not depend on it as long as the above conditions are fulfilled.} This requires the allowed sources to be all localized inside the box with sufficiently short temporal support.
	
	\pagebreak
	We now follow the steps \eqref{Delta_order_1} to \eqref{final_2nd_order} in order to derive an expression for the reflection propagator [using the same decomposition as in \eqref{G_decomposition}]:
	\begin{multline}\label{eq:G_R_pert}
		\!\!\!\!\! G^{\mathcal{R}}_{xy}= \mathrm{i} \! \int_0^\infty \!\!\!\!\mathrm  d z^\perp \!\! \int_k  \frac{ g_I^{\mu\nu} k_\mu k_\nu}{2 \left(k_\perp + \mathrm{i} \bar \epsilon \right)}  
		e^{-\mathrm{i} k_\perp (x^\perp + y^\perp) + 2\mathrm{i}  z^\perp(k_\perp + \mathrm{i} \bar \epsilon)}  \,  \\
		- \frac{\mathrm{i}}{2}\int_k  \left[ \frac{\sqrt{g_>^{\perp \perp}}}{\sqrt{f(R)}}-1\right]  e^{-\mathrm{i} k_\perp (x^\perp + y^\perp)} \, .
	\end{multline}
	
\begin{figure}
		\centering
		\includegraphics{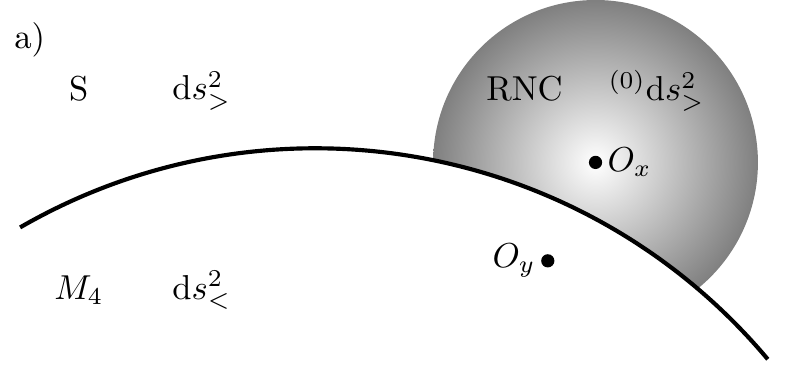}
		\includegraphics{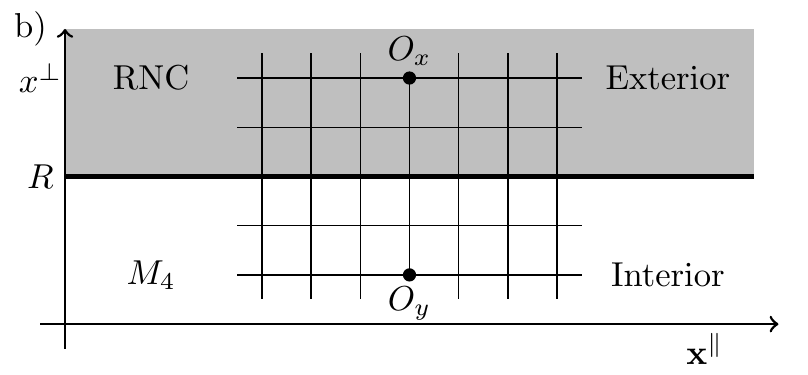}
		\caption{\label{collapseRNC} a) Spherical suspended shell with a RNC patch constructed at $O_x$ with a Schwarzschild background in the exterior and Minkowski spacetime in the interior with shifted origin at $O_y$. b) Zoomed-in situation with the shell as a boundary at $R$ with negligible curvature analogous to the two half-space system of the optical system in Fig.~\ref{figHalfSp}.}
	\end{figure}

	The spatial dependence of the interacting metric in~\eqref{eq:G_R_pert} needs to be specified for the $z^\perp$ integration.  In the next section, we will straightforwardly substitute \eqref{g_I_2}. Here instead, we will use a RNC construction anchored outside the shell, as shown in Fig.~\ref{collapseRNC}a). RNC approximate the spacetime locally around a reference point $r=r_\star$ and are introduced in detail in Appendix~\ref{appRNC}. At leading order, the corresponding metric is 
	\begin{multline}
		{}^{(0)}\mathrm d s_>^2 
		=- f(r_\star)\, \mathrm d t_S^2 + f^{-1}(r_\star) \,\mathrm d r^2 + r^2 \mathrm{d}\Omega_2 \, .
		\label{eq:RNCinSchS}
	\end{multline}
	If we were to use RNC, this metric simply becomes a Minkowski spacetime as explained in Appendix~\ref{appRNC}. Taking the RNC or the Schwarzschild coordinates in \eqref{eq:RNCinSchS} would correspond to a noncontinuous coordinate-slicing across the shell. Instead, we will use \eqref{insideTimeOutside} (for $\dot R = 0$) alongside \eqref{x_coord}, which leads to
	\begin{subequations}
		\label{RNC_cover}
		\begin{align}
			\mathrm ds^2_< &= - (\mathrm d x^t)^2  +(\mathrm d x^\perp)^2 + (\mathrm d \vec x^\parallel)^2\,,\label{insMink}\\
			\!\! {}^{(0)}\mathrm ds^2_> &= -\frac{ f(r_\star)}{f(R)}   ( \mathrm d x^t)^2  + \frac{1}{f(r_\star)} (\mathrm d x^\perp)^2 + (\mathrm d \vec x^\parallel)^2 \label{RNCinMink} \, ,
		\end{align}
	\end{subequations}
	where now the same coordinates are used in the interior ($x^\perp < 0$) and in the exterior ($x^\perp >0$).

	Crucially, the interior and exterior coordinate patches are not continuous over the shell at $x^\perp = 0$, but are a bipartite cover of the underlying smooth manifold, with the discontinuity containing information about the exterior geometry.\footnote{A simple but intuitive analogy is provided by a continuously differentiable function approximated by a piecewise constant but noncontinuous function. In our case, we approximate the manifold locally using only two ``pieces''.} In principle, this cover can be extended to any number of RNC patches, provided they all overlap. The size of each patch, i.e.~the spacetime volume within which it is a good approximation of the underlying manifold, is discussed in the Appendix~\ref{appRNC}. If a more accurate description of the external geometry is desired, more RNC patches can be glued together and/or higher order contributions to the RNC expansion can be considered, increasing the size of the individual patches.
	
	\pagebreak
	This RNC construction amounts to replacing $g_I(z)$ and $g_>(z)$ in \eqref{eq:G_R_pert} by $g_I(r_\star)$ and $g_>(r_\star)$, i.e., the exterior metric is evaluated at the reference point and thus loses its spatial dependence. As a result, $g_>^{\perp \perp}|_R = f(r_\star) \neq f(R)$ and the boundary contribution in the second row of \eqref{eq:G_R_pert} is nonvanishing. Moreover, the $z^\perp$ integration becomes trivial as it only depends on the plane waves. We obtain for the reflection propagator (evaluated inside)
	\begin{align}
		G^{\mathcal{R}}_{xy} =  -\int_k \, {\rm e}^{\mathrm{i} k_\perp (x^\perp + y^\perp)} \mathcal{R}(\vec{k}) \, ,
	\end{align}
	together with the reflection coefficient
	\begin{align}
		\! \! \mathcal{R}(\vec{k}) = \frac{\lambda \bar x_\star}{4}   \left[1+\tan^2(\alpha)\right] \,.
		\label{pertReflLead}
	\end{align}
	Recall that the smallness parameters $\lambda = r_g/R \ll 1$, ${\bar x_\star = x_\star/R \ll 1}$ with ${x_\star = r_\star-R}$ and the angle of incidence is ${\tan^2(\alpha) = \vec{k}_\parallel^2 / k_\perp^2}$. We see that the reflectance increases as the shell approaches the horizon formation. Compared to the optical example, the exterior curvature acts like a medium with radially decreasing susceptibility. This optical analogy is also supported by the observation that $\mathcal{R}$ increases with the angle of incidence. Physically, this makes sense since we are insensitive to curvature effects on very small scales. The previous condition that $x^\perp/R \ll 1$ (and $|\vec{x}^\parallel|/R \ll 1$) is the condition $(R k_\perp) \gg 1$ (and $(R k_\parallel) \gg 1$) in momentum space. The above expression is thus valid for sufficiently large frequencies with $\omega_k R > \omega_k r_g \gg 1 $ and hence complementary to the gray-body calculation, which makes a statement about the low-frequency range with $\omega_k r_g \ll 1 $~\cite{Page:1976df}.
	
	\pagebreak
	
	\subsection{Nonperturbative approach}\label{sec:non_pert}
	
	We have seen that the perturbative approach is not valid close to horizon formation since there $\lambda \rightarrow 1$. Therefore, we now consider the nonperturbative matching introduced in Sec.~\ref{sec:toy_np} to derive the reflection and transmission propagators relating both sides of the shell.
	
	As before, we work in the system depicted in fig.~\ref{collapseRNC}b) and use the RNC construction \eqref{RNCinMink}. In contrast to the previous section, $\lambda$ is not assumed to be small. Placing a source in the interior, we make an ansatz for the respective propagators inside ($x^\perp < 0$) and outside ($x^\perp > 0$) the shell,
	\begin{subequations}
		\begin{align}
			G^<_{xy} &=  \int_k \left[{\rm e}^{\mathrm{i} k_\perp (x^\perp - y^\perp)}  -  {\rm e}^{-\mathrm{i} k_\perp (x^\perp + y^\perp)} \mathcal{R}(\vec k)\right]\,,  \label{RNCinsideProp}\\
			G^{>}_{xy} &= \int_k  {\rm e}^{\mathrm{i} (q_\perp x^\perp /\sqrt{f(r_\star)}- k_\perp y^\perp)} \,  \mathcal{T}(\vec k) \, , \label{RNCoutsideProp}
		\end{align}
	\end{subequations}
	where $x^t > y^t$ was assumed and the factor $1/\sqrt{f(r_\star)}$ in \eqref{RNCoutsideProp} was introduced for later convenience. Here, $G^<_{xy}$ is the flat space expression (\ref{reflectionScalar}) with $\varepsilon=1$ and $G^>_{xy}$ generalizes \eqref{transScalar} to the exterior shell geometry. The momentum~$q_\perp$ is then fixed by the requirement $\Box_x G^>_{xy} = 0$, explicitly
	\begin{align}
		\quad q_\perp (\vec k) =\mathrm{sgn}(k_\perp) \sqrt{ \left( \frac{f(R)}{f(r_\star)} - 1 \right) \vec k_\parallel^2 + \frac{f(R)}{f(r_\star)} k_\perp^2} \, .
		\label{momRelScalarRNC}
	\end{align} 
	For an arbitrarily large shell with $R \rightarrow \infty$ while keeping $r_\star>R$, this reduces to $q_\perp = k_\perp$ since the curvature difference between the interior and exterior geometry vanishes in this limit. Moreover, for $R \rightarrow r_g$ and $r_\star > R$, the right-hand side vanishes in the normal incidence scenario. In the optical system, this corresponded to a total internal reflection. Here it is the manifestation of the event horizon.
	
	The coefficients $\mathcal{R}$ and $\mathcal{T}$ follow from the covariant version of the matching conditions~\eqref{bndryCond} which are given by
	\begin{subequations}
		\label{bndryCondRNC}
		\begin{align}
			\lim_{x^{\perp} \nearrow \, R} G^{<}_{xy}&= \lim_{x^{\perp} \searrow \, R} G^{>}_{xy} \, ,  \\
			\lim_{x^{\perp} \nearrow \, R} n^\mu_<\nabla_\mu^x G^{<}_{xy} &= \lim_{x^{\perp} \searrow \, R} n^\mu_>\nabla_\mu^x G^{>}_{xy}  \,,
		\end{align}
	\end{subequations}
	where $n^\mu_<=(0,1,0,0)$ and $n^\mu_> = (0,\sqrt{f(r_\star)},0,0)$ are the interior and exterior normal vectors (\ref{normVec_suspended}).
	By substituting the Green functions \eqref{RNCinsideProp} and \eqref{RNCoutsideProp}, one obtains analogously to the optical case 
	\begin{align}
		1 - \mathcal{R}(\vec k) &= \mathcal{T}(\vec k) \, , \quad k_\perp + k_\perp \mathcal{R}(\vec k) = q_\perp \mathcal{T}(\vec k) \,,
	\end{align}
	which can be readily solved for $\mathcal{R}$ and $\mathcal{T}$,
	\begin{align}
		\mathcal{R}(\vec k)&= \frac{q_\perp-k_\perp}{k_\perp+q_\perp} \, , \qquad  \mathcal{T}(\vec k)= \frac{2k_\perp}{k_\perp+q_\perp} \, .
		\label{tHalfMRNC}
	\end{align}
	Following the procedure in Sec.~\ref{consEsec}, these coefficients obey charge conservation and recover the expression as in Eq.~\eqref{chargeDensCons} with the momentum $q_\perp$ defined in \eqref{momRelScalarRNC}.
	
	In Sec.~\ref{sec:curvePert}, the perturbative analysis was performed for a suspended shell with a radius sufficiently larger than~$r_g$ and a normal neighborhood with expansion point close to the shell, i.e. $\lambda = r_g/R \ll 1$ and $\bar x_\star = x_\star/R \ll 1$ with $x_\star = r_\star-R$. Expanding the reflection coefficient up to second order in these smallness parameters yields 
	\begin{align}
		\! \! \mathcal{R}(\vec{k}) = \frac{\lambda \bar x_\star}{4}   \left[1+\tan^2(\alpha)\right] \Big( 1 - \bar x_\star + \lambda  \Big)
		\,,
		\label{Rpert2nd}
	\end{align}
	{where we neglected terms of order $\lambda^3, \lambda \bar x_\star^3$ and $\lambda^2 \bar x_\star^2$.}
	The leading order terms indeed agree with \eqref{pertReflLead} obtained in the perturbative approach. We note that including the boundary term in Eq.~\eqref{fullGreenFpg} was crucial for finding this agreement.
	
	The reflection coefficient in \eqref{Rpert2nd} depends sensitively on the exact position of the expansion point $x_\star$ of the normal neighborhood. If it were chosen too small, the reflectance would be underestimated because the curvature in the exterior region would not be probed sufficiently. On the other hand, if $x_\star$ were too large, the reflection could be overestimated. We now propose a procedure  to find an optimal choice for $x_\star$:
	We determine a more accurate yet perturbative expression for a reflection propagator that is not reliant on a RNC construction. To that end, we evaluate \eqref{fullGreenFpg} using the interaction metric~\eqref{g_I_2} and match the result with the one using the RNC expansion which in turn fixes the expansion point $x_\star$. We derive the propagator at next-to-leading order by following Appendix~\ref{app2ndorder}. As we have to take into account the spatial dependence of the interaction metric, the computation changes slightly: the $z^\perp$ integration is nontrivial (but solvable), the boundary contribution in \eqref{eq:G_R_pert} vanishes as $g_>^{\perp \perp}|_R = f(R)$ and the derivative in \eqref{app1PropOr2} now also acts on the transmission coefficient. Taking this into account, the reflection coefficient amounts to
	\begin{align}
		\! \! \! \!  \mathcal{R}_S(\vec{k}) = -\frac{\lambda}{8} \Bigg( \frac{ \tan^2(\alpha)}{ R k_{\bot }} \Big( \mathrm{i} + \frac{1}{R k_\perp} + \mathrm{i} \lambda \Big) -{\lambda}{} +\frac{3 \mathrm{i}  \lambda}{2 R k_{\bot }} \Bigg),
		\label{reflPertOr2}
	\end{align}
	{where analogous to \eqref{Rpert2nd} we neglected terms of order~$\lambda^3, \lambda/(R k_\perp)^3$ and $\lambda^2/(R k_\perp)^2$.}
	
	Note that the reflection propagator with the RNC construction and the one without the RNC construction differ only in their reflection coefficient $\mathcal{R}$ and $\mathcal{R}_S$, defined in Eq.~\eqref{Rpert2nd} and Eq.~\eqref{reflPertOr2}, respectively. We match them for $\tan(\alpha) \ll 1$, which will be of particular interest to us later. Solving for the expansion point, then yields
	\begin{align}
		\bar x_\star = \frac{1}{4} \left( 2 - \lambda \right) \lambda + \frac{1}{2} \left( \lambda - 1 \right) \lambda \tan^2(\alpha) + \mathcal O\left( \lambda^3, \tan^4(\alpha) \right)\!.
		\label{expPoint}
	\end{align}
	
	In the next section, we take the nonperturbative reflection propagator with expansion point \eqref{expPoint} to investigate quantum field theoretical properties of the suspended shell system.

	\section{Applications}
	\label{sec:QFTan}
	
	Using the tools developed in the previous section, we can now study the quantum consistency of a thin-shell system. Specifically, we ask what happens in a communication experiment across the surface of a shell near the formation of a black-hole at $r_g$ and whether the (perturbative) probabilistic interpretation of QFT is preserved in such a system.
	
	\subsection{Communication experiment}

	The approach outlined in the previous section was nonperturbative in the interaction parameter $\lambda$ and thus has the advantage of also capturing the case where the shell is close (and slightly beyond) horizon formation at $R \approx r_g $. We can therefore use the reflection coefficient $\mathcal{R}$ as a diagnostic probe for a communication experiment around horizon formation. 
	
\begin{figure}
		\centering
		\includegraphics{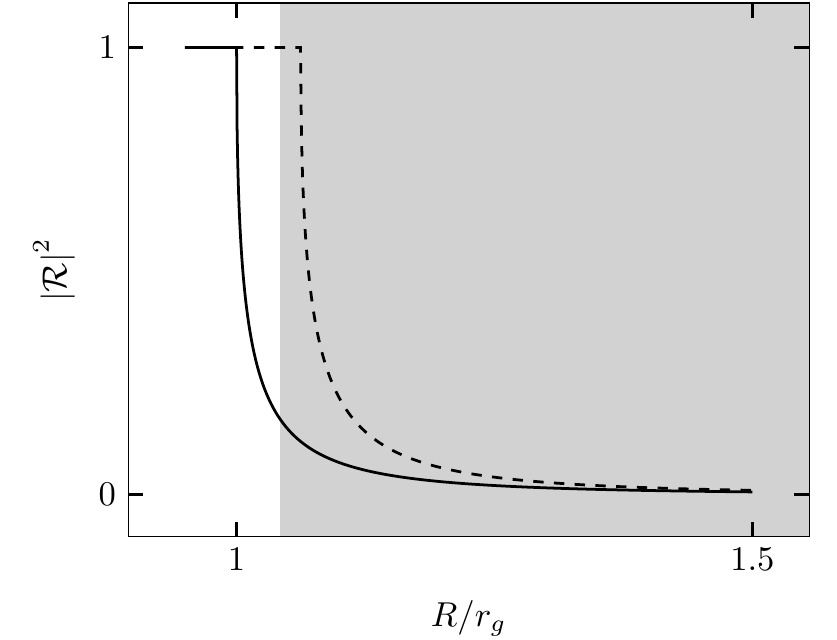}
		\caption{Reflectance from the inside of the shell according to Eq.~(\ref{tHalfMRNC}) for the suspended shell model with radius $R$ and RNC expansion in the exterior region, anchored at \eqref{expPoint}. The solid line is for a mode at normal incidence, i.e. $\alpha=0$, for which the reflectance reaches unity at $R=r_g$ (or $\lambda=1$ equivalently). The dashed line is for the case $\alpha=\pi/6$, for which total reflection occurs at a larger radius $R>r_g$. In the latter case, this occurs in a range where the fixed shell system can be trusted with $R>25r_g/24$, indicated by the shaded area.}
		\label{reflectFix}
	\end{figure}

	We consider the case where the detector is placed outside at $r_\star > r_g$ and the source inside the shell. We then study the system for different radii $R$. As we have seen before, any energy or charge transfer across the shell ceases when the transmittance $\mathrm{Re}(q_\perp) |\mathcal{T}|^2/k_\perp \to 0$. As a result of the continuity condition in~\eqref{chargeDensCons}, this also implies that the reflectance $|\mathcal{R}|^2 \to 1$. In other words, there is no communication between the shell's inside and outside possible. Due to \eqref{tHalfMRNC}, a sufficient condition for this to happen is  $q_\perp=0$. This defines a critical value of $\lambda$, which to leading order reads
	\begin{align}\label{lambda_c}
		\lambda_c = 1 - \bar{x}_\star \tan^2(\alpha) + \mathcal{O}(\bar{x}_\star^2)\,,
	\end{align} 
	where $\bar{x}_\star$ needs to be substituted with the generalized version of \eqref{expPoint} valid for $\alpha \neq 0$ (i.e.~away from the normal incidence limit). First we consider the case of normal incidence with $\alpha=0$ depicted as the solid line in Fig.~\ref{reflectFix}. Here the critical value evaluates to $\lambda_c =1$, which corresponds to the point of horizon crossing (as it should of course). This demonstrates that our optical approach reflects the causal structure of the underlying spacetime: Communication ceases at the latest at horizon crossing. For $\alpha \neq 0 $, this happens even earlier for values $\lambda < 1$ as demonstrated in \eqref{lambda_c} and by the dashed line in Fig.~\ref{reflectFix} for the choice $\alpha=\pi / 6$. This can be understood as follows. If the propagation is not in radial direction a particle is less efficient at evading the gravitational pull of the shell.  In the case of $\alpha=\pi/6$, the critical point $\lambda_c < 24/25$ and therefore falls within the range where the shell can be fully stabilized in terms of physical matter. 
	
	Moreover, the transition of the reflection coefficient to unity is continuous, as opposed to a jump at $r_g$, which would otherwise indicate a drastic change in the observables during the formation of a black-hole, which would contradict the equivalence principle.
	
	Comparing the reflection coefficient (\ref{tHalfMRNC}) with that obtained in the optical model (\ref{tHalfM}), we find that for the suspended shell the analog of total internal reflection in optics occurs once $\lambda < \lambda_c$. This corresponds to a situation where the momentum $q_\perp$ in (\ref{momRelScalarRNC}) turns imaginary, which in turn leads to an exponential damping of the modes outside the shell. In contrast to the optical scenario, this is not due to different susceptibilities, but to the curvature in the exterior of the shell.
	
	As discussed in Sec.~\ref{subsub_shellGeo}, a shell suspended too close to $r_g$ is unphysical. The consideration for $R\leq 25 r_g/24$ is therefore of more formal interest. A similar study where the collapse is maximally slowed down but horizon formation not avoided is left for future work. What we can do instead is to consistently investigate the regime in which the shell is close to forming a black-hole. Our simple communication experiment then illustrates that the boundary propagator approach is fully compatible with the conventional expectation obtained by studying the causal structure of a shell spacetime.
	
	\subsection{Interior vacuum persistence amplitude}
	
	\label{sec_vpa}
	
	As a first means of studying the quantum consistency of a suspended shell background, we use the vacuum persistence amplitude as a diagnostic tool. It indicates by how much the vacuum state is destabilized in the presence of an external source $J$. This calculation serves as a direct application of the propagator found in the previous section. We will use the local vacuum $| 0 \rangle$ associated with an inertial observer within the shell. To make this construction explicit we use the Minkowski line element~\eqref{insMink} inside the shell to construct the local quantum field, 
	\begin{align}
		\phi(x) = \int \frac{\mathrm d^3 k}{\sqrt{(2\pi)^3 2 \omega_k}} \left( {e^{i k_\mu x^\mu}} a_{\vec k} + e^{-i k_\mu x^\mu} a_{\vec k}^\dagger \right)  \, ,
		\label{qfield}
	\end{align}
	where $k_\mu x^\mu = -\omega_k x^t + k_\perp x^\perp + \vec k_\parallel \vec x^\parallel$, with creation operator $a_{\vec k}^\dagger$ and annihilation operator $a_{\vec k}$ defining the local vacuum through $a_{\vec k} |0 \rangle = 0$.
	We start with an initial Minkowski vacuum state $| 0_i \rangle = | 0 \rangle$ and calculate the probability with which it evolves into the final vacuum state $| 0_f \rangle = |0 \rangle$ while the external source $J$ is turned on and off again. The transition amplitude is \cite{Schwinger:1953zza}
	\begin{align}
		{\langle {0_f} | 0_i \rangle_J^0} = \mathrm{ e}^{ -\frac{{\mathrm{i}}}{2} \int\mathrm{d} \mu_x \mathrm d \mu_y \mathrm  \, J_x \Delta_{xy} J_y } \, ,
		\label{vpa}
	\end{align}
	where we normalized to the amplitude in the absence of the source $\langle {0_f} | 0_i \rangle_0$, i.e.~$\langle {0_f} | 0_i \rangle_J^0 := \langle {0_f} | 0_i \rangle_J/ \langle {0_f} | 0_i \rangle_0$. Since we place the external source $J$ inside the shell, Eq. (\ref{vpa}) encodes only nontrivial information about the external geometry through the propagator $\Delta$. An external source $J$ radiates and thus occupies the system with particles that were not present in the initial state. As a result, the transition probability decreases, which is a measure of the amount of particle production, explicitly
	\begin{align}
		|{\langle {0_f} | 0_i \rangle_J^0}|^2 \leq 1 \, .
		\label{VPAsm1}
	\end{align}
	On the other hand, $|{\langle {0_f} | 0_i \rangle_J^0}|^2>1$ would signal an inconsistency and be at odds with the probabilistic interpretation of QFT. Therefore, the calculation of $|{\langle {0_f} | 0_i \rangle_J^0}|^2$ is an important consistency check. In particular, a violation of (\ref{VPAsm1}) would call into question the validity of the semiclassical approximation, according to which the background is treated purely classically.
	
	To investigate the transition amplitude (\ref{vpa}) for the suspended shell system, we use the nonperturbative reflection propagator (\ref{RNCinsideProp}) with coefficient \eqref{tHalfMRNC}. Since this propagator was found by applying different approximation methods, we have to choose an external source that takes all these approximations into account consistently. As discussed in detail in the Appendix~\ref{appRNC}, a specific choice of an external source that meets these criteria is one that is spatially pointlike, located at $\vec x_J$, but smeared in time through a Gaussian profile with standard deviation $\sigma_t$, i.e.~$J_x \propto \delta^{(3)}(\vec x-{\vec x}_J) \exp\left[-(x^t)^2/ (2 \sigma_t^2)\right]$. In momentum space this leads to
	\begin{align}
		J(\vec k) = \frac{1}{\sqrt{2\pi}}  \mathrm{e}^{\mathrm i \vec k \vec x_J} \mathrm{e}^{ - \frac{(\omega_k- \langle \omega_k \rangle)^2}{2 } \sigma_t^2  } \, ,
		\label{gaussSorc}
	\end{align}
	where we introduced the mean $\langle \omega_k \rangle$, which needs to be sufficiently large to ensure that infrared divergences can be ignored. $\sigma_t$ and $\langle \omega_k \rangle$ are both constrained by the validity of the approximations used. {Specifically, we choose $\sigma_t /r_g  = 0.08 \ll 1$ and $\langle \omega_k \rangle  r_g = 4 r_g/(\sigma_t ) \gg 1$ (which is also compatible with the RNC expansion as argued in Appendix~\ref{appRNC}).}
	
	\begin{figure}
		\centering
		\includegraphics{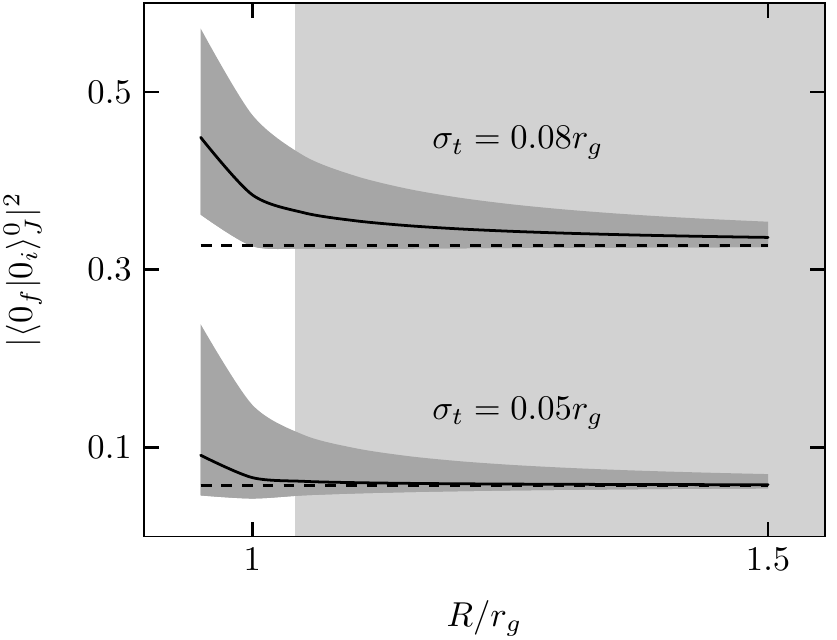}
		\caption{Vacuum persistence amplitude (\ref{vpa}) for the Minkowski vacuum for a temporally smeared-out point source \eqref{gaussSorc} inside a shell of constant radius $R$. The two dashed lines are the contributions, which are insensitive to the shell and its external geometry and serve as a reference for the Minkowski contributions. The solid lines and the gray bands take into account the reflection from the exterior geometry. We choose the distance between the source and the shell $0.005 r_g$ for the central solid lines and $0$ for the upper and $0.01 r_g$ for the lower boundary of the dark shaded regions. The RNC expansion is anchored at \eqref{expPoint} for $\alpha=0$ and the source parameters are $\sigma_t = 0.08 r_g$ and $\langle \omega_k \rangle = 4/\sigma_t$ for the upper plot and $\sigma_t = 0.05 r_g$ for the lower plot. All parameters are chosen in accordance with the validity discussion in the Appendix~\ref{appRNC}. The shaded area in the background indicates the radii $R > 25 r_g/24$ for which the shell can be stabilized with respect to standard matter.}
		\label{plotVPA}
	\end{figure}
	
	Taking these considerations into account, we can compute the probability $|{\langle {0_f} | 0_i \rangle_J^0}|^2$ with Eq. (\ref{vpa}) for an external source located in the interior close to the shell. To be precise, for our numerical example we consider a source distance to the shell between $0$ and $0.01 r_g$. The result as a function of the shell radius $R/r_g$ (or $1/\lambda$ equivalently) and the distance between the shell and the source is shown in Fig.~\ref{plotVPA}. We can make several immediate observations: First, the probability in absence of the shell, shown as the dashed lines, tends towards unity for large $\sigma_t$. This corresponds to a source which is turned on and off adiabatically in Minkowski space, for which no particle production is expected. 
	Second, the deviation from the Minkowski result becomes significant (but not singular) as $R=r_g$ is approached. We interpret this as a change in the vacuum structure induced by the onset of black-hole formation. Third, the relevance of the reflection contribution decreases the farther the source is placed away from the surface. This reflects the spatial falloff of the propagator, making the effect of the boundary less pronounced when it is farther away. Fourth, the vacuum persistence amplitude is smooth and always smaller than unity. Therefore, in this setup, the probabilistic interpretation of QFT is preserved and no pathologies arise. However, what we can also observe in Fig.~\ref{plotVPA} is that the longer the external source is turned on, i.e.~the larger $\sigma_t$ is, the larger the vacuum persistence amplitude becomes. Importantly, this effect is more pronounced the closer we are to black-hole formation. This raises the question as to whether this enhancement leads to any pathological behavior in the limit where $\sigma_t$ is comparable with the lifetime of the black-hole~\cite{Page:1976df}.  We believe that this long-term behavior can be studied using Fermi normal coordinates (FNC) which a priori do not suffer from the same temporal restriction \cite{Hoegl:2020hif} as the RNC construction.
	
	\section{Discussion and Outlook}
	\label{sec:conclusion}
	
	We started out studying propagation within and across media of different susceptibility. This extensive analysis allowed us to recover known results from geometrical optics in the language of Green's functions, validating our approach. In particular, we presented two independent computations of the Feynman propagator, one being perturbative in the difference of susceptibilities and the other one nonperturbative.
	
	In a next step, we set up our suspended shell model. To that end, we calculated the surface pressure needed to stabilize a thin shell at radius $R$ by using Israel's junction conditions and inferred a lower bound on $R$ by demanding that the shell matter fulfills the dominant and strong energy condition. We then approximated the geometry in terms of two distinct RNC patches providing a local covering of the interior and exterior geometry sufficiently close to the shell. We could then formally map this geometry to our previously discussed two optically active media. Again, we used both the perturbative and nonperturbative approach to derive an explicit expression for the reflective and transmissive part of the Green's function.
	
	As a first sanity check of our computation, we confirmed that no on-shell signal can leave the shell interior once it crosses its own horizon, which manifests itself through a reflection coefficient that approaches unity. For larger radii, where the shell can be stabilized in terms of physical matter, signals can leave the shell as the transmission is turned on and the reflection turned off continuously. This is caused by the decrease in curvature outside the shell (which we capture at leading order by our two-patch covering).
	
	After these preliminary considerations, we calculated the vacuum persistence amplitude inside the shell. It provides a measure of the vacuum stability in the presence of an external source. We found that the amplitude as compared to the pure Minkowski case, i.e.~without a shell,  receives an enhancement  when the source is located close enough to the shell. This effect becomes more pronounced if the source is turned on for a longer time and/or when the shell moves closer to horizon crossing. Despite this effect, the amplitude never exceeds unity in agreement with a unitary time evolution. However, there is an important caveat as we could only probe the geometry on microscopic timescales due to the limited temporal extend of the exterior Minkowski patch. In other words, the enhancement effect we observed has the potential to turn into a pathology on long enough timescales. We believe that we can get around this limitation by using a FNC construction related to an orbital observer outside the shell. While we want to study this generalization in our future work, the present article lies out the technical and methodological groundwork needed to do so.
	
	This will enable us to study two main questions in our future work:
	\begin{itemize}
		\item Does the persistence amplitude of the suspended shell vacuum respect the unitarity bound in Eq.~\eqref{VPAsm1} on arbitrary long timescales?
		\item Is horizon formation in a retarded shell model, where the collapse is maximally slowed down, semiclassically consistent?
	\end{itemize}
	
	Our approach admits many other sophistications and extensions which merit further exploration. For example, calculating the expectation value of the energy-momentum tensor in the shell vacuum will provide a complementary picture closer to previous analyses in the literature (see for example \cite{Akhmedov:2015xwa}). Moreover, turning on self-interactions (rather than external sources) is another way of testing the robustness of our results under model alterations. Improving the approximation of the exterior geometry is another priority. This can happen in two straightforward ways by either including  more RNC patches or going to higher order in the RNC expansion, which would both extend the covering of the exterior manifold. Finally, there is the question of how Hawking radiation manifests itself in this framework. At this stage, we only speculate that it can be related to the finite penetration depth we observed in the case of total reflection when $R < r_g$ and the radial momentum turns imaginary leading to a nonvanishing yet exponentially damped support of the mode functions outside the shell.
	
	\appendix*
	\section{}
	\label{sec: appendix}
	\renewcommand{\thesubsection}{\Alph{subsection}}
	
	\subsection{Second order perturbative calculation}
	\label{app2ndorder}
	Here, we evaluate {in detail} the first and second order term in the expansion \eqref{perturbative_expansion}. We begin with the first order contribution
	\begin{align}\label{Delta_order_1}
		\Delta_{xy}^{(1)} := \lambda_S \int \limits_S \mathrm d \mu_{z} \, \partial_{z^t} \Delta^\varepsilon_{x z}  \partial_{z^t} \Delta^\varepsilon_{z y}\,.
	\end{align}
	After inserting the Feynman propagators we have
	\begin{align}
		\lambda_S \int \limits_S \mathrm d \mu_{z} \int \frac{\mathrm d^4 k}{(2\pi)^4} \frac{\mathrm d^4 q}{(2\pi)^4}  \frac{k_0 q_0 \, e^{\mathrm{i}q(x-z)}e^{\mathrm{i}k(z-y)}}{\left(\eta^{\mu\nu}_\varepsilon q_\mu q_\nu -\mathrm{i} \bar \epsilon)(\eta^{\mu\nu}_\varepsilon k_\mu k_\nu -\mathrm{i} \bar \epsilon\right)}\, ,
	\end{align}
	where we introduced the $\mathrm{i} \bar \epsilon$ prescription with $\bar \epsilon$ in order to distinguish it from the susceptibility $\varepsilon$. The limit $\bar \epsilon \rightarrow 0$ outside the momentum integrals  is understood.
	Performing the $z^t$ and  $\vec z^\parallel$ integration yields delta distributions, which in turn collapse the corresponding $q_0$ and $\vec q_\parallel$ integration,
	\begin{multline}\label{zt_and_z_p}
		\lambda_S \int_0^\infty \mathrm  d z^\perp \int \frac{\mathrm d^4 k}{(2\pi)^4} \frac{\mathrm d q_\perp}{2\pi}  
		e^{\mathrm{i} \vec k_\parallel(\vec x^\parallel-\vec y^\parallel)}  \\
		\times \frac{k_0^2 \, e^{-\mathrm{i} k_0(x^t-y^t)}e^{-\mathrm{i}  z^\perp (q_\perp -k_\perp )}e^{\mathrm{i} q_\perp x^\perp - \mathrm{i} k_\perp y^\perp}}{(-\varepsilon k_0^2 + \vec k_\parallel^2 + q_\perp^2 -\mathrm{i} \bar \epsilon)(-\varepsilon k_0^2 + \vec k_\parallel^2 + k_\perp^2 -\mathrm{i} \bar \epsilon)}\,  \,.
	\end{multline}
	There are in total four poles in the complex $k_0$ plane. We assume $x^t > y^t$ and thus only pick up two of them in the lower half-plane,
	\begin{multline}\label{k0_int}
		\mathrm{i} \lambda_S  \int_0^\infty \mathrm  d z^\perp \int \frac{\mathrm d^3 k}{(2\pi)^3} \frac{\mathrm d q_\perp}{2\pi  } e^{\mathrm{i} \vec k_\parallel(\vec x^\parallel-\vec y^\parallel)} e^{\mathrm{i} q_\perp (x^\perp-z^\perp)}\\
		\times \frac{ \omega^\varepsilon_k {e^{-\mathrm{i} \omega_k^\varepsilon (x^t-y^t)}} - \omega^\varepsilon_q {e^{-\mathrm{i} \omega_q^\varepsilon (x^t-y^t)}} }{2\varepsilon(k_\perp - q_\perp) (k_\perp + q_\perp)}  e^{\mathrm{i}  k_\perp (z^\perp-y^\perp)}\,  \,,
	\end{multline}
	with definition $\omega_q^\varepsilon := \sqrt{\vec k_\parallel^2 + q_\perp^2}/\sqrt{\varepsilon}$. Next, we apply for convenience the Sokhotski-Plemelj theorem, $\mathcal{P}\left[1/(k_\perp \pm q_\perp)\right] = 1/(k_\perp \pm q_\perp + \mathrm{i} \bar \epsilon) + \mathrm{i} \pi \delta(k_\perp \pm q_\perp)$, with the limit $\bar \epsilon \rightarrow 0$ understood. The delta distributions yield vanishing contributions, since the two terms in the numerator cancel when $k_\perp = \mp q_\perp$; explicitly,
	\begin{multline}\label{first_order_step}
		\mathrm{i}  \lambda_S \int_0^\infty \mathrm  d z^\perp \int \frac{\mathrm d^3 k}{(2\pi)^3} \frac{\mathrm d q_\perp}{2 \pi} e^{\mathrm{i} \vec k_\parallel(\vec x^\parallel-\vec y^\parallel)}e^{\mathrm{i}  q_\perp (x^\perp-z^\perp)}\\
		\times\frac{  \omega^\varepsilon_k {e^{-\mathrm{i} \omega_k^\varepsilon (x^t-y^t)}} -\omega^\varepsilon_q {e^{-\mathrm{i} \omega_q^\varepsilon (x^t-y^t)}}}{2 \varepsilon (k_\perp-q_\perp + \mathrm{i} \bar \epsilon)(k_\perp + q_\perp + \mathrm{i} \bar \epsilon)}  e^{\mathrm{i} k_\perp (z^\perp-y^\perp)} \,  \,.
	\end{multline}
	The term proportional to $\omega_q^\varepsilon$ introduces the poles ${k_\perp = q_\perp - \mathrm{i} \bar \epsilon}$ and $k_\perp = -q_\perp - \mathrm{i} \bar \epsilon$ in the lower complex $k_\perp$ half-plane. This term evaluates to zero since $z^\perp > 0 > y^\perp $. The term proportional to $\omega_k^\varepsilon$, on the other hand, yields a nonvanishing contribution closing the $q_\perp$ integration contour in the lower (for $x^\perp < z^\perp$) or upper half-plane (for $x^\perp > z^\perp$),
	\begin{multline}\label{first_order_prop}
		\mathrm{i} \lambda_S  \int_0^\infty \mathrm  d z^\perp \int_k^{\varepsilon}  \frac{ (\omega^\varepsilon_k)^2}{2 \left(k_\perp + \mathrm{i} \bar \epsilon \right)}  \Big[ \theta_{x^\perp z^\perp} e^{\mathrm{i} k_\perp (x^\perp - y^\perp)} \\
		- \theta_{z^\perp x^\perp} e^{-\mathrm{i} k_\perp (x^\perp + y^\perp) + 2\mathrm{i}  z^\perp(k_\perp + \mathrm{i} \bar \epsilon)}  \Big]\,  \,.
	\end{multline}
	Finally, the $z^\perp$ integration results in
	\begin{multline}\label{first_order_result}
		\Delta_{xy}^{(1)}\Big|_{x^t>y^t} =- \int_k^\varepsilon \frac{ \bar \omega}{4 } \Big[ \theta_{x^\perp} e^{\mathrm{i} k_\perp (x^\perp  - y^\perp)} \left(1 -2\mathrm{i}  x^\perp k_\perp  \right) \\
		+ \theta_{-x^\perp} e^{-\mathrm{i} k_\perp (x^\perp + y^\perp)}  \Big]\,  \,,
	\end{multline}
	which yields together with $\theta_{x^\perp} \Delta^{\varepsilon}_{xy}$ the expressions in \eqref{reflExpand} and \eqref{transmission_prop_2n_order} evaluated at order $\lambda_S$.
	
	We are now ready to compute the second order term in \eqref{perturbative_expansion},
	\begin{align}\label{sec_order_prop}
		{\Delta_{xy}^{(2)} :=} \lambda_S^2  \int \limits_S \mathrm d \mu_{z_1}  d \mu_{z_2} \, \partial_{z_1^t}\Delta^\varepsilon_{x z_1} \partial_{z_1^t}  \partial_{z_2^t}\Delta^\varepsilon_{z_1 z_2} \partial_{z_2^t} \Delta^\varepsilon_{{z_2} y}\;.
	\end{align}
	
	In order to simplify calculations, we express the linear transmission propagator $\Delta^{\mathcal{T}(1)}_{xy} := \Delta_{xy}^{(1)}|_{x^\perp > 0}$ as a four momentum integral, 
	\begin{align}
		\Delta^{\mathcal{T}(1)}_{xy} =  \int \frac{\mathrm d^4 k}{(2\pi)^4}    \frac{{e^{\mathrm{i}  k(x-y)}}}{\left(\eta^{\mu\nu}_{\varepsilon}  k_\mu   k_\nu - \mathrm{i} \bar \epsilon \right)} X(k_0, k_\perp, x^\perp)\,  \,,
	\end{align}
	with 
	\begin{align}\label{X}
		X(k_0, k_\perp, x^\perp) := -\lambda_s k_0^2 \, \frac{   1 -2 \mathrm{i}  x^\perp \left(k_\perp + \mathrm{i} \bar  \epsilon \right) }{4 \left(k_\perp + \mathrm{i} \bar  \epsilon \right)^2} \,.
	\end{align}
	It is easy to check that for $x^t>y^t$ this recovers the term in \eqref{first_order_result} proportional to $\theta_{x^\perp}$ when performing the $k_0$ integration. There is also a second contribution, only present for $x^t < y^t$ and hence not contributing in \eqref{k0_int}, which is now crucial for the calculation of the second order.
	We can now solve higher orders iteratively. Substituting back into \eqref{sec_order_prop}  and keeping in mind that $z_1^\perp>0$ yields
	
	\begin{align} 
		\Delta_{xy}^{(2)} = \lambda_S \int \limits_S \mathrm d \mu_{z} \, \partial_{z^t} \Delta^\varepsilon_{x z} \mathcal  \partial_{z^t} \Delta_{zy}^{\mathcal T(1)}\,.
		\label{app1PropOr2}
	\end{align}
	
	The next steps can be performed like in \eqref{zt_and_z_p} and \eqref{k0_int}, providing us with a generalized version of \eqref{first_order_step} (again assuming $x^t>y^t$),
	\begin{multline}
		\mathrm{i}  \lambda_S \int_0^\infty \mathrm \!\! d z^\perp\!\! \int \frac{\mathrm d^3 k}{(2\pi)^3} \frac{\mathrm d q_\perp}{2 \pi} e^{\mathrm{i} \vec k_\parallel(\vec x^\parallel-\vec y^\parallel)}e^{\mathrm{i}  q_\perp (x^\perp-z^\perp)}\\ 
		\times \frac{  \omega^\varepsilon_k {e^{-\mathrm{i} \omega_k^\varepsilon (x^t-y^t)}} X(\omega_k^\varepsilon,k_\perp ,z^\perp) - (\omega^\varepsilon_k \rightarrow \omega^\varepsilon_q) }{2  \varepsilon (k_\perp-q_\perp + \mathrm{i} \bar  \epsilon)(k_\perp + q_\perp + \mathrm{i} \bar \epsilon)} e^{\mathrm{i} k_\perp (z^\perp-y^\perp)}  .
	\end{multline}
	The term proportional to $\omega_q^\varepsilon$ introduces three poles in the lower half of the complex $k_\perp$ plane. They yield a vanishing contribution because the integration contour needs to be closed in the upper half-plane as $z^\perp > y^\perp$. The term proportional to $\omega_k^\varepsilon$, by contrast, has the same pole structure as the corresponding term in \eqref{first_order_step}, resulting in a nonvanishing contribution:
	\begin{multline}\label{first_order_prop_2nd}
		\!\!\!\mathrm{i} \lambda_S  \int_0^\infty \mathrm  d z^\perp \int_k^{\varepsilon}  \frac{ (\omega^\varepsilon_k)^2 X(\omega^\varepsilon_k, k_\perp ,z^\perp)}{2 \left(k_\perp + {\mathrm{i} \bar \epsilon} \right)}  \Big[ \theta_{x^\perp z^\perp} e^{\mathrm{i} k_\perp (x^\perp - y^\perp)} \\
		- \theta_{z^\perp x^\perp} e^{-\mathrm{i} k_\perp (x^\perp + y^\perp) + 2\mathrm{i}  z^\perp(k_\perp + \mathrm{i} \bar \epsilon)}  \Big] \, .
	\end{multline}
	Substituting \eqref{X} allows us to perform the $z^\perp$ integration.  After separating terms proportional to $\theta(x^\perp)$ and $\theta(-x^\perp)$, we have
	\begin{multline}\label{final_2nd_order}
		\Delta_{xy}^{(2)}\Big|_{x^t>y^t} =\\
		\int_k^{\varepsilon}   \frac{ \bar \omega^2}{8  } \Big[ \theta_{x^\perp} e^{\mathrm{i} k_\perp (x^\perp  - y^\perp)} \Big( 1 - 2 \mathrm{i} k_\perp x^\perp -  (x^\perp)^2 k_\perp^2 \Big)\\
		+\theta_{-x^\perp} e^{-\mathrm{i} k_\perp (x^\perp  + y^\perp)} \Big]\,,
	\end{multline}
	{which results in the second order contribution to \eqref{reflExpand} and \eqref{transmission_prop_2n_order}.} Higher orders can then be calculated iteratively by again separating out the transmission part from \eqref{final_2nd_order} and using it as an input for the subsequent order.

	\subsection{Junction conditions}
	\label{subsec: appendix Israel}
	Here, we work out the junction conditions for a collapsing shell.
	In Sec.~$a$ we apply Israel's formalism in terms of extrinsic curvature~\cite{Israel:1966rt}. In Sec.~$b$ we take a distributional approach.
	
	\subsubsection{Extrinsic curvature}
	
	We consider a collapsing shell with trajectory $R(\tau)$. The geometry  outside the shell is parametrized through Schwarzschild coordinates $x^\mu_> =~(t_S, r, \vartheta, \varphi)$ with line element \eqref{SchSLine} and inside through spherical Minkowski coordinates $x^\mu_< = (x^t, r, \vartheta, \varphi)$. Denoting the coordinates on the shell as $\tilde x^{\tilde \mu} = (\tau, \vartheta, \varphi)$, the embedding of the shell is given by $F_>^\mu = (t_S(\tau), R(\tau), \vartheta, \varphi)$ and  $F_<^\mu = (x^t(\tau), R(\tau), \vartheta, \varphi)$. The induced metric on the shell is then given as the pullback
	\begin{align}
		\tilde g_{\tilde \mu \tilde \nu}=  \frac{\partial {F}_\circ^{\alpha\vphantom{\beta}} }{\partial \tilde  x^{\tilde \mu}} \, \frac{\partial {F}_\circ^\beta}{\partial \tilde x^{\tilde \nu}} \, g_{\alpha\beta}^\circ = \text{diag}\left(-1, R^2(\tau), R^2(\tau) \sin^2(\vartheta) \right)\, ,
	\end{align}
	where $\circ \in \{>, <\}$ distinguishes the cases where the metric is induced from the outside or inside, respectively. The equality of both expressions then ensures the continuity of the line element. It corresponds to the first junction condition and relates the time inside with the time outside:
	\begin{align}
		\frac{\mathrm d t_S}{\mathrm d x^t} = \frac{\sqrt{f(R) + \dot R^2}}{f(R) \sqrt{1 + \dot R^2 }} \, ,
		\label{timeRelSchMink}
	\end{align}
	where $\dot R = \mathrm d R/ \mathrm d \tau$. 
	
	The second junction condition states that a jump J across the shell in the extrinsic curvature $K$ only occurs if there is an energy-momentum tensor $\tilde T^{\tilde \mu}_{\tilde \nu}$ induced on the boundary, explicitly
	\begin{align}
		\text{J}\left(\tilde K^{\tilde \mu}_{\tilde \nu}\right) - \delta^{\tilde \mu}_{\tilde \nu} \text{J}\left(K\right) = -8 \pi \tilde T^{\tilde \mu}_{\tilde \nu} \,.
		\label{einsteinFE}
	\end{align}
	Unlike a dust cloud, the shell has a distributional character, and we expect a nonvanishing jump of $K$ across the shell. The extrinsic curvature is
	\begin{align}
		K_{\mu\nu} = \frac{1}{2} \mathcal L_n h_{\mu\nu} = h_\mu^\alpha h_\nu^\beta \nabla_\alpha n_\beta \, ,
	\end{align}
	where we introduced the tensor $h_{\mu\nu} = g_{\mu\nu} - n_\mu n_\nu$ and the shell normal vector $n_\mu \propto \partial_\mu (r- R)$ obeying $g^{\mu\nu} n_\mu n_\nu = 1$. Evaluated at the surface of the shell the normal vector inferred from the outside geometry becomes
	\begin{align}
		n^>_\mu(R) =\left(- \dot R , \frac{\sqrt{f(R)+\dot R^2}}{f(R)}, 0, 0\right)  ,
		\label{normVec}
	\end{align}
	while the normal vector from inside $n^<_\mu$ follows by formally replacing $f(R)$ with $1$ in the expression of~$n^>_\mu$. Since the normal vector can also be expressed as $n^>_\mu =~(- \dot R, \dot t_S,0 ,0)$, it is evident that it is orthogonal to the 4-velocity of the shell $u^\mu_> = (\dot t_S, \dot R, 0, 0)$.
	
	As an ansatz for the shell's energy-momentum tensor we take a perfect fluid with energy density $\rho$ and pressure $p$. Calculating the extrinsic curvature and substituting it back into the junction condition \eqref{einsteinFE} then uniquely determines for every shell trajectory $R(t)$ the required energy density and pressure. Specifically, if we evaluate the $\tau \tau$ component of \eqref{einsteinFE}, we find the energy density~(\ref{shellPre}). The pressure equation can be derived from the spatial components, or by computing the conservation equation $\tilde \nabla^{\tilde \nu} \tilde T^{\tilde \mu}_{\tilde \nu} = 0$. Choosing the latter, the corresponding pressure is expressed in terms of the energy density as~(\ref{shellPre}) in agreement with \cite{doi:10.1142/S0218271817501589}.
	
	\subsubsection{Distributional description}
	Alternatively, spacetimes with boundaries can conveniently be described distributional. In order to recall the concepts, we choose a simplified example first. Consider a subspace of the real numbers $\Omega \subset \mathbb R$ and name the function of interest $f \in \mathcal C^1(\Omega)$ together with test functions $\varphi \in \mathcal C_0^\infty(\Omega)$ with support $\text{supp}(\varphi) \subset \mathcal K = [-c ,c] \subset \Omega$. We denote the convolution as
	\begin{align}
		[f'] \varphi = \int_{\mathbb R} \mathrm d \tau f'(\tau) \varphi(\tau) = -\int_{\mathcal K} \mathrm d \tau f(\tau) \varphi'(\tau) \, ,
	\end{align}
	where we integrated by parts in the second step and used the support properties of $\varphi$. Let us now take out one single point $\tau_{\mathcal B} \in (-c,c)$ with $f \in C^1(\mathbb  R \backslash \{\tau_{\mathcal B}\})$ and introduce a jump $j_{\mathcal B}$ in $f$ around this point with $\lim_{\lambda \rightarrow 0} \left( f(\tau_{\mathcal B} + \lambda) - f(\tau_{\mathcal B} - \lambda)\right) = j_{\mathcal B}$. In this case we start with
	\begin{align}
		[f]' \varphi := - \lim_{\lambda \rightarrow 0} \left( \int_{-c}^{\tau_{\mathcal B}-\lambda}  +  \int_{{\tau_{\mathcal B}+\lambda}}^c  \right) \mathrm d \tau f(\tau) \varphi'(\tau) \, .
	\end{align}
	Since $\varphi$ is continuous across $\tau_{\mathcal B}$ and vanishes at $\pm c$ the boundary terms that arise when integrating by parts yield $j_{\mathcal B} \varphi (\tau_{\mathcal B})$. Thus, we find 
	\begin{align}
		[f]' \varphi = [f'] \varphi + j_{\mathcal B} \varphi(\tau_{\mathcal B}) \, .
	\end{align}

	In a 4-dim spacetime manifold $\mathcal M$, a hypersurface can be a 3-dim submanifold that is either timelike, spacelike or lightlike. Let $\zeta= (x^\alpha)=(x^0,x^1,x^2,x^3)$ denote a coordinate system in spacetime with appropriate domain. A particular timelike or spacelike hypersurface ${\mathcal B}$ can be endowed with a coordinate representation by putting a restriction on the coordinates $f(\zeta)=0$, or by giving parametric equations of the form $\zeta=\zeta(\sigma)$ where $\sigma$ denotes a coordinate system in ${\mathcal B}$. Locally, the value of $f$ changes only in the direction perpendicular to ${\mathcal B}$ and thus $\text{grad}(f) \perp {\mathcal B}$. A unit normal $n$ can be introduced by $g(n,n)=\kappa$ with $\kappa=+1$ for ${\mathcal B}$ timelike and $\kappa=-1$ for ${\mathcal B}$ spacelike. Demanding $\nabla_n f > 0$ we can thus write $n = \kappa \, \text{grad}(f) / | g(\mathrm df,\mathrm df) |^{1/2}$.
	
	Let $\sigma=(y^a)=(y^1,y^2,y^3)$ denote a particular coordinate system in ${\mathcal B}$. Then as usual $e_a := \partial_a =: e^\alpha_a \partial_\alpha$. The line element in $\sigma$ is $\mathrm d s^2|_{\mathcal B} = g_{\alpha\beta} \mathrm d x^\alpha \mathrm d x^\beta =: h_{ab} \mathrm d y^a \mathrm d y^b$ with induced metric $h$ of the hypersurface. A measure on ${\mathcal B}$ can be introduced as follows: In the specified coordinate neighborhood $\mathrm d \Sigma := | \text{det}(h)|^{1/2} \mathrm d^3 y$ is called a surface element. The combination $n \mathrm d \Sigma$ is the directed surface element that by definition points in the direction of increasing $f$.
	
	Let $\mathcal R$ be an open submanifold of $\mathcal M$, then a congruence of geodesics $C(\mathcal R)$ is a family of curves such that through each point $p\in \mathcal R$ there passes precisely one curve in this family. These geodesics are parametrized with $\tau$ such that $\forall p \in \mathcal R\, \exists ! \, \gamma \in C(\mathcal R) : \exists \, \tau_p \in \mathbb R, \tau_p \geq \tau_B : \gamma(\tau_p) = p$. 
	
	Now consider a hypersurface $\mathcal B$ that partitions a manifold $\mathcal M$ into two regions $\mathcal R^>$ and $\mathcal R^<$ with metric tensors $g^>$ and $g^<$, respectively. The corresponding generalized metric tensor on $(\mathcal M, \mathcal B)$ is given by $g = \chi_{>} g^> + \chi_{<} g^<$ with support function $\chi_{>}(p) = 1 : p \in \mathcal R^>$ and $\chi_{>}(p) = 0$ otherwise. Also, we defined the complement $\chi_< = \chi_>^C$. On this partitioned manifold we define a generalized Levi-Civita connection $D$ and generalized tensor fields. With these we define the generalized Riemannian curvature tensor of $(\mathcal M, \mathcal B)$ through the usual formula
	\begin{align}
		R(X,Y) Z = D_{[X,Y]} Z - [D_X,D_Y] Z\, ,
	\end{align}
	with generalized vector fields $X, Y$ and $Z$ on $(\mathcal M, \mathcal B)$. The crucial difference to the usual Riemann tensor is that
	\begin{multline}
		D_X D_Y Z = \chi_> (D_X D_Y Z)^> + \chi_< (D_X D_Y Z)^<  \\
		+ X(\chi_>) \text{J}_{\mathcal B}(D_Y Z)\, ,
	\end{multline}
	with jump function $\text{J}_{\mathcal B}(D_Y Z) = (D_Y Z)^> - (D_Y Z )^< $ on~$\mathcal B$. As a consequence, $R \not = \xi^> R^> + \chi^< R^<$; instead, it includes jump terms too:
	\begin{multline}
		R(X,Y)Z = \chi_> (R(X,Y)Z)^> + \chi_< (R(X,Y)Z)^<  \\
		- \left[ X(\chi_>) \text{J}_{\mathcal B}(D_Y Z) - Y(\chi_<) \text{J}_{\mathcal B}(D_X Z) \right] \,.
	\end{multline}
	
	Suppose that $\mathcal B$ is pierced by a congruence of geodesics orthogonal to it with $n$ denoting the corresponding normal vector field on $\mathcal B$. Then with the Riemann tensor we can substitute the Ricci tensor Ric and derive the jump function
	\begin{align}
		\text{J}_{\mathcal B}[\text{Ric}(X,Y)] = \text{J}_{\mathcal B}\left[ g( D_{P^{\parallel} X} n, Y )\right] \delta_{\tau_{\mathcal B}} \, ,
		\label{jumpRic}
	\end{align}
	where $P^{\parallel}$ projects to the parallel part of $X$ to $\mathcal B$.
	Therefore, extending the Einstein equation to hold for the corresponding distribution valued tensors a jump function for the Ricci tensor and scalar can only be caused by a distributional valued energy-momentum tensor on $\mathcal B$.
	
	As examples we take the suspended shell in Schwarzschild coordinates and the collapsing shell with velocity $\mathrm d R/ \mathrm d \tau =-\sqrt{1-f(R)}$ in Painlev\'e-Gullstrand coordinates. Computing the jump functions of the Ricci tensor (\ref{jumpRic}) for the former with normal vector at the shell $n_S^\mu(R)=(0,\sqrt{f(R)},0,0)$ results in
	\begin{align}
		\text{J}_{\mathcal B}[ g( D_{\partial^{\tau}} n_S, \partial_{\tau} )] \delta_{\tau_{\mathcal B}} &= \frac{r_g}{2 R^2 \sqrt{f(R)}} \, ,\nonumber \\
		\text{J}_{\mathcal B}[ g( D_{\partial^\vartheta} n_S, \partial_\vartheta )] \delta_{\tau_{\mathcal B}} &= \frac{  \sqrt{f(R)}-1}{R}\, ,
	\end{align}
	where the $\vartheta\vartheta$ component in the second line equals the $\varphi\varphi$ component. Substituting these jumps together with their traces in the Einstein field equations results in a shell as perfect fluid with energy density and pressure \eqref{fixShellRhoPre}.
	
	Taking the collapsing shell in Painlev\'e-Gullstrand coordinates the normal vector becomes $n_P^\mu(R) = (0,1,0,0)$. Then the jump for the Ricci tensor (\ref{jumpRic}) becomes for the collapsing shell
	\begin{align}
		\text{J}_{\mathcal B}[ g( D_{\partial^{\tau}} n_P, \partial_{\tau} )] \delta_{\tau_{\mathcal B}} &= \frac{r_g}{2 R^2 \sqrt{f_+(R)}} \, ,\nonumber \\
		\text{J}_{\mathcal B}[ g( D_{\partial^\vartheta} n_P, \partial_\vartheta )] \delta_{\tau_{\mathcal B}} &= \frac{ 1 - \sqrt{f_+(R)}}{R}\, .
	\end{align}
	Inserting this into the Einstein field equations results in a collapsing shell with energy density and pressure \eqref{PGShellRhoPre}.

	\subsection{Normal Coordinate Systems}
	\label{appRNC}
	
	According to the principle of relativity, the gravitational and inertial masses are equal. As a consequence, an observer cannot distinguish whether they are accelerating or subjected to a homogeneous gravitational field. Therefore, for an arbitrarily curved background, locally in a small neighborhood where the gravitational field is sufficiently homogeneous, the metric is flat in the coordinates associated with a free-falling observer at rest. For larger neighborhoods where the Minkowski metric is insufficient, one can Taylor expand the inhomogeneity of the gravitational field, leading to correction terms. 
	
	One coordinate manifestation of this procedure is the RNC, which can be derived as follows. Given the coordinate velocity $v^a$ of the observer in a background metric $g$ expressed in the global coordinates $x^a$, we find locally the vielbein $e^a_\mu$ satisfying the relations $e^a_t = v^a$ and $g_{ab} e^a_\alpha e^b_\beta = \eta_{\alpha\beta}$. With this vielbein, we can now construct the RNC metric series in the usual coordinates $\xi^\alpha$ \cite{Petrov}:
	\begin{equation}
		g_{\alpha\beta}(\xi) = \eta_{\alpha\beta} + \frac 1 3 R_{\alpha\mu\nu\beta} \xi^\mu \xi^\nu + \mathcal{O}\left(\xi^3\right)\,,
		\label{RNCmet}
	\end{equation}
	with the Riemann tensor evaluated at the origin $\xi^\alpha = 0$. If one chooses the spacetime points for an experiment sufficiently close to the origin, the higher order terms in $\xi^\alpha$ are negligible and it is sufficient to use the Minkowski patch. On the other hand, if one moves away from the origin the error increases. In the following, we will therefore provide an error analysis for the systems we investigate in Sec.~\ref{sec:QFTan}.
	
	As already pointed out, the truncation of the metric of a normal coordinate system at an adiabatic order restricts the spacetime region in which this metric can be trusted. The size and structure of this region has been studied in detail in \cite{Hoegl:2020hif}. We are interested in considering only the Minkowski contribution, i.e.~the first term in (\ref{RNCmet}). The minimum Schwarzschild radius that can be described in this patch around the expansion radius~$r_\star$ is given by Eq. (30) in \cite{Hoegl:2020hif} and reads
	\begin{equation}
		\label{rmin}
		r_{\text{min}} = r_\star \left(1-\frac{3}{\sqrt{2}}\sqrt{\delta}+\frac{15}{4}\delta\right)^{\frac 2 3} \, ,
	\end{equation}
	where $\delta$ is the maximal error resulting from neglecting the terms of adiabatic order 2 and 3. For the desired expansion point in \eqref{expPoint}, the maximal radius for the expansion point is $r_\star = 5/4 R$ for $R\rightarrow r_g$. If we require that the Minkowski patch reaches the suspended shell of radius $r_\text{min}=R$, we obtain the upper bound $\delta \approx 5 \times 10^{-2}$ with Eq.~\eqref{rmin}, which constitutes an acceptable error. As explained in the main text, the mismatch that causes this error is entirely due to the overestimation of the metric in the half of the RNC patch between the expansion point and the shell, and is compensated to some extent by the other half of the patch where the metric is underestimated.
	
	Starting from a RNC patch, one can obtain a propagator that respects the metric up to a desired adiabatic order by performing an expansion for large momenta \cite{bunchparker}. This propagator can be conveniently calculated with (\ref{fullGreenFpg}) by considering the interaction Lagrangian as the difference between the kinetic term of the field in the RNC background and the kinetic term in Minkowski. The resulting series is not trustworthy for finite adiabatic order if the momenta are too small or equivalently the wavelength is too large, i.e. on the order of the curvature length scale. This can be taken care of by introducing an infrared energy cutoff $\omega_\text{IR}$. If we require that the error due to neglecting the contribution of the next-to-leading order only affects the result by a maximum of 1 percent near $r_\star$, the cutoff value must be chosen to be $\omega_\text{IR} \approx 4 \sqrt{r_g /2}/(r_\star)^{3/2}$.
	
	Because of this cutoff, any infrared-sensitive observable is strongly dependent on the exact value of $\omega_\text{IR}$. In systems with external sources $J$, however, we can choose $J$ such that the contribution of infrared physics is negligible. To be specific, if we assume a Gaussian dependence in energy space for the source as in (\ref{gaussSorc}), we must choose the standard deviation $\sigma_t$ and the mean $\langle \omega_k \rangle$ accordingly. Moreover, for a given choice of $\sigma_t$, we must also ensure that the temporal validity of the system under consideration is large enough: The temporal validity of a RNC patch according to Eq. (29) of \cite{Hoegl:2020hif} depends on the mass of the black-hole. Taking $\delta = 10^{-2}$ as before, a Minkowski patch anchored at $r_\star$ holds for $|x^t| \lesssim t_\text{max}(R) = 0.16 \, r_\star^{3/2}\sqrt{f(R)}/[r_g f(r_\star)]$. In Sec.~\ref{sec_vpa} we use the same source (\ref{gaussSorc}) for different values of $R$. Since $t_\text{max}$ increases monotonically for $R>r_g$, we choose the most restrictive setting for which the suspended shell is viable, i.e. $R \rightarrow 25 r_g/24$. Therefore, taking $r_\star$ as in \eqref{expPoint}, we obtain $\sigma_t = t_\text{max}(25 r_g/24)/2 \approx 0.08 r_g$ and $\langle \omega_k \rangle = 4/\sigma_t$ to account for the cutoff $\omega_\text{IR}$. For astrophysical black-holes with a mass range from 1 to $10^{11}$ solar masses, the maximum temporal validity ranges from $t_\text{max} = 1 \,\mu$s to $t_\text{max} = 40\,$h in SI units.
	
	Once the time restriction of the external source is fixed, only its spatial restriction remains to be determined. In Sec.~\ref{sec_vpa} we choose a point source located inside the shell at $\vec x_J$. Since we have approximated the surface of the shell to be flat, the source needs to be placed sufficiently close to the surface. Assuming that the error in approximating the shell surface as flat is of the order of $10^{-2}$, this restricts the distance between the shell and the source $R- x_J^\perp < 0.1 R$ in agreement with the choice made in our VPA analysis in Sec.~\ref{sec_vpa}.
	
	\begin{acknowledgments}
		It is a great pleasure to thank Cecilia Giavoni, Marc Schneider and Martin S.~Sloth for inspiring and valuable comments and suggestions.
	\end{acknowledgments}
	
	\bibliography{library.bib}

\begin{thebibliography}{33}%
\makeatletter
\providecommand \@ifxundefined [1]{%
 \@ifx{#1\undefined}
}%
\providecommand \@ifnum [1]{%
 \ifnum #1\expandafter \@firstoftwo
 \else \expandafter \@secondoftwo
 \fi
}%
\providecommand \@ifx [1]{%
 \ifx #1\expandafter \@firstoftwo
 \else \expandafter \@secondoftwo
 \fi
}%
\providecommand \natexlab [1]{#1}%
\providecommand \enquote  [1]{``#1''}%
\providecommand \bibnamefont  [1]{#1}%
\providecommand \bibfnamefont [1]{#1}%
\providecommand \citenamefont [1]{#1}%
\providecommand \href@noop [0]{\@secondoftwo}%
\providecommand \href [0]{\begingroup \@sanitize@url \@href}%
\providecommand \@href[1]{\@@startlink{#1}\@@href}%
\providecommand \@@href[1]{\endgroup#1\@@endlink}%
\providecommand \@sanitize@url [0]{\catcode `\\12\catcode `\$12\catcode
  `\&12\catcode `\#12\catcode `\^12\catcode `\_12\catcode `\%12\relax}%
\providecommand \@@startlink[1]{}%
\providecommand \@@endlink[0]{}%
\providecommand \url  [0]{\begingroup\@sanitize@url \@url }%
\providecommand \@url [1]{\endgroup\@href {#1}{\urlprefix }}%
\providecommand \urlprefix  [0]{URL }%
\providecommand \Eprint [0]{\href }%
\providecommand \doibase [0]{https://doi.org/}%
\providecommand \selectlanguage [0]{\@gobble}%
\providecommand \bibinfo  [0]{\@secondoftwo}%
\providecommand \bibfield  [0]{\@secondoftwo}%
\providecommand \translation [1]{[#1]}%
\providecommand \BibitemOpen [0]{}%
\providecommand \bibitemStop [0]{}%
\providecommand \bibitemNoStop [0]{.\EOS\space}%
\providecommand \EOS [0]{\spacefactor3000\relax}%
\providecommand \BibitemShut  [1]{\csname bibitem#1\endcsname}%
\let\auto@bib@innerbib\@empty
%</preamble>
\bibitem [{\citenamefont {Abbott}\ \emph {et~al.}(2016)\citenamefont {Abbott}
  \emph {et~al.}}]{Abbott:2016blz}%
  \BibitemOpen
  \bibfield  {author} {\bibinfo {author} {\bibfnamefont {B.}~\bibnamefont
  {Abbott}} \emph {et~al.} (\bibinfo {collaboration} {LIGO Scientific,
  Virgo}),\ }\href {https://doi.org/10.1103/PhysRevLett.116.061102} {\bibfield
  {journal} {\bibinfo  {journal} {Phys. Rev. Lett.}\ }\textbf {\bibinfo
  {volume} {116}},\ \bibinfo {pages} {061102} (\bibinfo {year} {2016})},\
  \Eprint {https://arxiv.org/abs/1602.03837} {arXiv:1602.03837 [gr-qc]}
  \BibitemShut {NoStop}%
\bibitem [{\citenamefont {Akiyama}\ \emph {et~al.}(2019)\citenamefont {Akiyama}
  \emph {et~al.}}]{Akiyama:2019cqa}%
  \BibitemOpen
  \bibfield  {author} {\bibinfo {author} {\bibfnamefont {K.}~\bibnamefont
  {Akiyama}} \emph {et~al.} (\bibinfo {collaboration} {Event Horizon
  Telescope}),\ }\href {https://doi.org/10.3847/2041-8213/ab0ec7} {\bibfield
  {journal} {\bibinfo  {journal} {Astrophys. J.}\ }\textbf {\bibinfo {volume}
  {875}},\ \bibinfo {pages} {L1} (\bibinfo {year} {2019})},\ \Eprint
  {https://arxiv.org/abs/1906.11238} {arXiv:1906.11238 [astro-ph.GA]}
  \BibitemShut {NoStop}%
\bibitem [{\citenamefont {Hawking}(1976)}]{Hawking:1976ra}%
  \BibitemOpen
  \bibfield  {author} {\bibinfo {author} {\bibfnamefont {S.}~\bibnamefont
  {Hawking}},\ }\href {https://doi.org/10.1103/PhysRevD.14.2460} {\bibfield
  {journal} {\bibinfo  {journal} {Phys. Rev. D}\ }\textbf {\bibinfo {volume}
  {14}},\ \bibinfo {pages} {2460} (\bibinfo {year} {1976})}\BibitemShut
  {NoStop}%
\bibitem [{\citenamefont {Page}(1976)}]{Page:1976df}%
  \BibitemOpen
  \bibfield  {author} {\bibinfo {author} {\bibfnamefont {D.~N.}\ \bibnamefont
  {Page}},\ }\href {https://doi.org/10.1103/PhysRevD.13.198} {\bibfield
  {journal} {\bibinfo  {journal} {Phys. Rev. D}\ }\textbf {\bibinfo {volume}
  {13}},\ \bibinfo {pages} {198} (\bibinfo {year} {1976})}\BibitemShut
  {NoStop}%
\bibitem [{\citenamefont {Giddings}(2007)}]{Giddings:2007ie}%
  \BibitemOpen
  \bibfield  {author} {\bibinfo {author} {\bibfnamefont {S.~B.}\ \bibnamefont
  {Giddings}},\ }\href {https://doi.org/10.1103/PhysRevD.76.064027} {\bibfield
  {journal} {\bibinfo  {journal} {Phys. Rev. D}\ }\textbf {\bibinfo {volume}
  {76}},\ \bibinfo {pages} {064027} (\bibinfo {year} {2007})},\ \Eprint
  {https://arxiv.org/abs/hep-th/0703116} {arXiv:hep-th/0703116} \BibitemShut
  {NoStop}%
\bibitem [{\citenamefont {Akhmedov}\ \emph {et~al.}(2016)\citenamefont
  {Akhmedov}, \citenamefont {Godazgar},\ and\ \citenamefont
  {Popov}}]{Akhmedov:2015xwa}%
  \BibitemOpen
  \bibfield  {author} {\bibinfo {author} {\bibfnamefont {E.~T.}\ \bibnamefont
  {Akhmedov}}, \bibinfo {author} {\bibfnamefont {H.}~\bibnamefont {Godazgar}},\
  and\ \bibinfo {author} {\bibfnamefont {F.~K.}\ \bibnamefont {Popov}},\ }\href
  {https://doi.org/10.1103/PhysRevD.93.024029} {\bibfield  {journal} {\bibinfo
  {journal} {Phys. Rev. D}\ }\textbf {\bibinfo {volume} {93}},\ \bibinfo
  {pages} {024029} (\bibinfo {year} {2016})},\ \Eprint
  {https://arxiv.org/abs/1508.07500} {arXiv:1508.07500 [hep-th]} \BibitemShut
  {NoStop}%
\bibitem [{\citenamefont {Burgess}\ \emph {et~al.}(2018)\citenamefont
  {Burgess}, \citenamefont {Hainge}, \citenamefont {Kaplanek},\ and\
  \citenamefont {Rummel}}]{Burgess:2018sou}%
  \BibitemOpen
  \bibfield  {author} {\bibinfo {author} {\bibfnamefont {C.}~\bibnamefont
  {Burgess}}, \bibinfo {author} {\bibfnamefont {J.}~\bibnamefont {Hainge}},
  \bibinfo {author} {\bibfnamefont {G.}~\bibnamefont {Kaplanek}},\ and\
  \bibinfo {author} {\bibfnamefont {M.}~\bibnamefont {Rummel}},\ }\href
  {https://doi.org/10.1007/JHEP10(2018)122} {\bibfield  {journal} {\bibinfo
  {journal} {JHEP}\ }\textbf {\bibinfo {volume} {10}},\ \bibinfo {pages}
  {122}},\ \Eprint {https://arxiv.org/abs/1806.11415} {arXiv:1806.11415
  [hep-th]} \BibitemShut {NoStop}%
\bibitem [{\citenamefont {Lunin}\ and\ \citenamefont
  {Mathur}(2002{\natexlab{a}})}]{Lunin:2001jy}%
  \BibitemOpen
  \bibfield  {author} {\bibinfo {author} {\bibfnamefont {O.}~\bibnamefont
  {Lunin}}\ and\ \bibinfo {author} {\bibfnamefont {S.~D.}\ \bibnamefont
  {Mathur}},\ }\href {https://doi.org/10.1016/S0550-3213(01)00620-4} {\bibfield
   {journal} {\bibinfo  {journal} {Nucl. Phys. B}\ }\textbf {\bibinfo {volume}
  {623}},\ \bibinfo {pages} {342} (\bibinfo {year} {2002}{\natexlab{a}})},\
  \Eprint {https://arxiv.org/abs/hep-th/0109154} {arXiv:hep-th/0109154}
  \BibitemShut {NoStop}%
\bibitem [{\citenamefont {Mazur}\ and\ \citenamefont
  {Mottola}(2001)}]{Mazur:2001fv}%
  \BibitemOpen
  \bibfield  {author} {\bibinfo {author} {\bibfnamefont {P.~O.}\ \bibnamefont
  {Mazur}}\ and\ \bibinfo {author} {\bibfnamefont {E.}~\bibnamefont
  {Mottola}},\ }\href@noop {} {\  (\bibinfo {year} {2001})},\ \Eprint
  {https://arxiv.org/abs/gr-qc/0109035} {arXiv:gr-qc/0109035} \BibitemShut
  {NoStop}%
\bibitem [{\citenamefont {Lunin}\ and\ \citenamefont
  {Mathur}(2002{\natexlab{b}})}]{Lunin:2002qf}%
  \BibitemOpen
  \bibfield  {author} {\bibinfo {author} {\bibfnamefont {O.}~\bibnamefont
  {Lunin}}\ and\ \bibinfo {author} {\bibfnamefont {S.~D.}\ \bibnamefont
  {Mathur}},\ }\href {https://doi.org/10.1103/PhysRevLett.88.211303} {\bibfield
   {journal} {\bibinfo  {journal} {Phys. Rev. Lett.}\ }\textbf {\bibinfo
  {volume} {88}},\ \bibinfo {pages} {211303} (\bibinfo {year}
  {2002}{\natexlab{b}})},\ \Eprint {https://arxiv.org/abs/hep-th/0202072}
  {arXiv:hep-th/0202072} \BibitemShut {NoStop}%
\bibitem [{\citenamefont {Dvali}\ and\ \citenamefont
  {Gomez}(2013)}]{Dvali:2011aa}%
  \BibitemOpen
  \bibfield  {author} {\bibinfo {author} {\bibfnamefont {G.}~\bibnamefont
  {Dvali}}\ and\ \bibinfo {author} {\bibfnamefont {C.}~\bibnamefont {Gomez}},\
  }\href {https://doi.org/10.1002/prop.201300001} {\bibfield  {journal}
  {\bibinfo  {journal} {Fortsch. Phys.}\ }\textbf {\bibinfo {volume} {61}},\
  \bibinfo {pages} {742} (\bibinfo {year} {2013})},\ \Eprint
  {https://arxiv.org/abs/1112.3359} {arXiv:1112.3359 [hep-th]} \BibitemShut
  {NoStop}%
\bibitem [{\citenamefont {Dvali}\ and\ \citenamefont
  {Gomez}(2012)}]{Dvali:2012wq}%
  \BibitemOpen
  \bibfield  {author} {\bibinfo {author} {\bibfnamefont {G.}~\bibnamefont
  {Dvali}}\ and\ \bibinfo {author} {\bibfnamefont {C.}~\bibnamefont {Gomez}},\
  }\href@noop {} {\  (\bibinfo {year} {2012})},\ \Eprint
  {https://arxiv.org/abs/1212.0765} {arXiv:1212.0765 [hep-th]} \BibitemShut
  {NoStop}%
\bibitem [{\citenamefont {Almheiri}\ \emph {et~al.}(2013)\citenamefont
  {Almheiri}, \citenamefont {Marolf}, \citenamefont {Polchinski},\ and\
  \citenamefont {Sully}}]{Almheiri:2012rt}%
  \BibitemOpen
  \bibfield  {author} {\bibinfo {author} {\bibfnamefont {A.}~\bibnamefont
  {Almheiri}}, \bibinfo {author} {\bibfnamefont {D.}~\bibnamefont {Marolf}},
  \bibinfo {author} {\bibfnamefont {J.}~\bibnamefont {Polchinski}},\ and\
  \bibinfo {author} {\bibfnamefont {J.}~\bibnamefont {Sully}},\ }\href
  {https://doi.org/10.1007/JHEP02(2013)062} {\bibfield  {journal} {\bibinfo
  {journal} {JHEP}\ }\textbf {\bibinfo {volume} {02}},\ \bibinfo {pages}
  {062}},\ \Eprint {https://arxiv.org/abs/1207.3123} {arXiv:1207.3123 [hep-th]}
  \BibitemShut {NoStop}%
\bibitem [{\citenamefont {Hofmann}\ and\ \citenamefont
  {Rug}(2016)}]{Hofmann:2014jya}%
  \BibitemOpen
  \bibfield  {author} {\bibinfo {author} {\bibfnamefont {S.}~\bibnamefont
  {Hofmann}}\ and\ \bibinfo {author} {\bibfnamefont {T.}~\bibnamefont {Rug}},\
  }\href {https://doi.org/10.1016/j.nuclphysb.2015.11.008} {\bibfield
  {journal} {\bibinfo  {journal} {Nucl. Phys. B}\ }\textbf {\bibinfo {volume}
  {902}},\ \bibinfo {pages} {302} (\bibinfo {year} {2016})},\ \Eprint
  {https://arxiv.org/abs/1403.3224} {arXiv:1403.3224 [hep-th]} \BibitemShut
  {NoStop}%
\bibitem [{\citenamefont {Gruending}\ \emph {et~al.}(2015)\citenamefont
  {Gruending}, \citenamefont {Hofmann}, \citenamefont {M\"uller},\ and\
  \citenamefont {Rug}}]{Gruending:2014rja}%
  \BibitemOpen
  \bibfield  {author} {\bibinfo {author} {\bibfnamefont {L.}~\bibnamefont
  {Gruending}}, \bibinfo {author} {\bibfnamefont {S.}~\bibnamefont {Hofmann}},
  \bibinfo {author} {\bibfnamefont {S.}~\bibnamefont {M\"uller}},\ and\
  \bibinfo {author} {\bibfnamefont {T.}~\bibnamefont {Rug}},\ }\href
  {https://doi.org/10.1007/JHEP05(2015)047} {\bibfield  {journal} {\bibinfo
  {journal} {JHEP}\ }\textbf {\bibinfo {volume} {05}},\ \bibinfo {pages}
  {047}},\ \Eprint {https://arxiv.org/abs/1407.1051} {arXiv:1407.1051 [hep-th]}
  \BibitemShut {NoStop}%
\bibitem [{\citenamefont {Bousso}(2020)}]{Bousso:2018bli}%
  \BibitemOpen
  \bibfield  {author} {\bibinfo {author} {\bibfnamefont {R.}~\bibnamefont
  {Bousso}},\ }\bibinfo {title} {{Black hole entropy and the Bekenstein
  bound}}\ (\bibinfo {year} {2020})\ pp.\ \bibinfo {pages} {139--158},\ \Eprint
  {https://arxiv.org/abs/1810.01880} {arXiv:1810.01880 [hep-th]} \BibitemShut
  {NoStop}%
\bibitem [{\citenamefont {Almheiri}\ \emph {et~al.}(2020)\citenamefont
  {Almheiri}, \citenamefont {Hartman}, \citenamefont {Maldacena}, \citenamefont
  {Shaghoulian},\ and\ \citenamefont {Tajdini}}]{Almheiri:2019qdq}%
  \BibitemOpen
  \bibfield  {author} {\bibinfo {author} {\bibfnamefont {A.}~\bibnamefont
  {Almheiri}}, \bibinfo {author} {\bibfnamefont {T.}~\bibnamefont {Hartman}},
  \bibinfo {author} {\bibfnamefont {J.}~\bibnamefont {Maldacena}}, \bibinfo
  {author} {\bibfnamefont {E.}~\bibnamefont {Shaghoulian}},\ and\ \bibinfo
  {author} {\bibfnamefont {A.}~\bibnamefont {Tajdini}},\ }\href
  {https://doi.org/10.1007/JHEP05(2020)013} {\bibfield  {journal} {\bibinfo
  {journal} {JHEP}\ }\textbf {\bibinfo {volume} {05}},\ \bibinfo {pages}
  {013}},\ \Eprint {https://arxiv.org/abs/1911.12333} {arXiv:1911.12333
  [hep-th]} \BibitemShut {NoStop}%
\bibitem [{\citenamefont {Penington}\ \emph {et~al.}(2019)\citenamefont
  {Penington}, \citenamefont {Shenker}, \citenamefont {Stanford},\ and\
  \citenamefont {Yang}}]{Penington:2019kki}%
  \BibitemOpen
  \bibfield  {author} {\bibinfo {author} {\bibfnamefont {G.}~\bibnamefont
  {Penington}}, \bibinfo {author} {\bibfnamefont {S.~H.}\ \bibnamefont
  {Shenker}}, \bibinfo {author} {\bibfnamefont {D.}~\bibnamefont {Stanford}},\
  and\ \bibinfo {author} {\bibfnamefont {Z.}~\bibnamefont {Yang}},\ }\href@noop
  {} {\  (\bibinfo {year} {2019})},\ \Eprint {https://arxiv.org/abs/1911.11977}
  {arXiv:1911.11977 [hep-th]} \BibitemShut {NoStop}%
\bibitem [{\citenamefont {Dvali}\ \emph {et~al.}(2020)\citenamefont {Dvali},
  \citenamefont {Eisemann}, \citenamefont {Michel},\ and\ \citenamefont
  {Zell}}]{Dvali:2020wft}%
  \BibitemOpen
  \bibfield  {author} {\bibinfo {author} {\bibfnamefont {G.}~\bibnamefont
  {Dvali}}, \bibinfo {author} {\bibfnamefont {L.}~\bibnamefont {Eisemann}},
  \bibinfo {author} {\bibfnamefont {M.}~\bibnamefont {Michel}},\ and\ \bibinfo
  {author} {\bibfnamefont {S.}~\bibnamefont {Zell}},\ }\href
  {https://doi.org/10.1103/PhysRevD.102.103523} {\bibfield  {journal} {\bibinfo
   {journal} {Phys. Rev. D}\ }\textbf {\bibinfo {volume} {102}},\ \bibinfo
  {pages} {103523} (\bibinfo {year} {2020})},\ \Eprint
  {https://arxiv.org/abs/2006.00011} {arXiv:2006.00011 [hep-th]} \BibitemShut
  {NoStop}%
\bibitem [{\citenamefont {Schwinger}(1953)}]{Schwinger:1953zza}%
  \BibitemOpen
  \bibfield  {author} {\bibinfo {author} {\bibfnamefont {J.}~\bibnamefont
  {Schwinger}},\ }\href {https://doi.org/10.1103/PhysRev.91.728} {\bibfield
  {journal} {\bibinfo  {journal} {Phys. Rev.}\ }\textbf {\bibinfo {volume}
  {91}},\ \bibinfo {pages} {728} (\bibinfo {year} {1953})}\BibitemShut
  {NoStop}%
\bibitem [{\citenamefont {Berczi}\ \emph {et~al.}(2021)\citenamefont {Berczi},
  \citenamefont {Saffin},\ and\ \citenamefont {Zhou}}]{Berczi:2020nqy}%
  \BibitemOpen
  \bibfield  {author} {\bibinfo {author} {\bibfnamefont {B.}~\bibnamefont
  {Berczi}}, \bibinfo {author} {\bibfnamefont {P.~M.}\ \bibnamefont {Saffin}},\
  and\ \bibinfo {author} {\bibfnamefont {S.-Y.}\ \bibnamefont {Zhou}},\ }\href
  {https://doi.org/10.1103/PhysRevD.104.L041703} {\bibfield  {journal}
  {\bibinfo  {journal} {Phys. Rev. D}\ }\textbf {\bibinfo {volume} {104}},\
  \bibinfo {pages} {L041703} (\bibinfo {year} {2021})},\ \Eprint
  {https://arxiv.org/abs/2010.10142} {arXiv:2010.10142 [gr-qc]} \BibitemShut
  {NoStop}%
\bibitem [{\citenamefont {Danielsson}\ \emph {et~al.}(2017)\citenamefont
  {Danielsson}, \citenamefont {Dibitetto},\ and\ \citenamefont
  {Giri}}]{Danielsson:2017riq}%
  \BibitemOpen
  \bibfield  {author} {\bibinfo {author} {\bibfnamefont {U.~H.}\ \bibnamefont
  {Danielsson}}, \bibinfo {author} {\bibfnamefont {G.}~\bibnamefont
  {Dibitetto}},\ and\ \bibinfo {author} {\bibfnamefont {S.}~\bibnamefont
  {Giri}},\ }\href {https://doi.org/10.1007/JHEP10(2017)171} {\bibfield
  {journal} {\bibinfo  {journal} {JHEP}\ }\textbf {\bibinfo {volume} {10}},\
  \bibinfo {pages} {171}},\ \Eprint {https://arxiv.org/abs/1705.10172}
  {arXiv:1705.10172 [hep-th]} \BibitemShut {NoStop}%
\bibitem [{\citenamefont {Hawking}(1975{\natexlab{a}})}]{Hawking:1974sw}%
  \BibitemOpen
  \bibfield  {author} {\bibinfo {author} {\bibfnamefont {S.}~\bibnamefont
  {Hawking}},\ }\href {https://doi.org/10.1007/BF02345020} {\bibfield
  {journal} {\bibinfo  {journal} {Commun. Math. Phys.}\ }\textbf {\bibinfo
  {volume} {43}},\ \bibinfo {pages} {199} (\bibinfo {year}
  {1975}{\natexlab{a}})},\ \bibinfo {note} {[Erratum: Commun.Math.Phys. 46, 206
  (1976)]}\BibitemShut {NoStop}%
\bibitem [{\citenamefont {Schwartz}(2014)}]{Schwartz:2013pla}%
  \BibitemOpen
  \bibfield  {author} {\bibinfo {author} {\bibfnamefont {M.~D.}\ \bibnamefont
  {Schwartz}},\ }\href@noop {} {\emph {\bibinfo {title} {{Quantum Field Theory
  and the Standard Model}}}}\ (\bibinfo  {publisher} {Cambridge University
  Press},\ \bibinfo {year} {2014})\BibitemShut {NoStop}%
\bibitem [{\citenamefont {Birkhoff}(1923)}]{ssbirkhoff}%
  \BibitemOpen
  \bibfield  {author} {\bibinfo {author} {\bibfnamefont {G.~D.}\ \bibnamefont
  {Birkhoff}},\ }\href {https://doi.org/10.4159/harvard.9780674734487} {\emph
  {\bibinfo {title} {{Relativity and Modern Physics}}}},\ Vol.\ \bibinfo
  {volume} {23008297}\ (\bibinfo  {publisher} {Harvard University Press},\
  \bibinfo {year} {1923})\ pp.\ \bibinfo {pages} {273--276}\BibitemShut
  {NoStop}%
\bibitem [{\citenamefont {Oppenheimer}\ and\ \citenamefont
  {Snyder}(1939)}]{PhysRev.56.455}%
  \BibitemOpen
  \bibfield  {author} {\bibinfo {author} {\bibfnamefont {J.~R.}\ \bibnamefont
  {Oppenheimer}}\ and\ \bibinfo {author} {\bibfnamefont {H.}~\bibnamefont
  {Snyder}},\ }\href {https://doi.org/10.1103/PhysRev.56.455} {\bibfield
  {journal} {\bibinfo  {journal} {Phys. Rev.}\ }\textbf {\bibinfo {volume}
  {56}},\ \bibinfo {pages} {455} (\bibinfo {year} {1939})}\BibitemShut
  {NoStop}%
\bibitem [{\citenamefont {Penrose}(1965)}]{PhysRevLett.14.57}%
  \BibitemOpen
  \bibfield  {author} {\bibinfo {author} {\bibfnamefont {R.}~\bibnamefont
  {Penrose}},\ }\href {https://doi.org/10.1103/PhysRevLett.14.57} {\bibfield
  {journal} {\bibinfo  {journal} {Phys. Rev. Lett.}\ }\textbf {\bibinfo
  {volume} {14}},\ \bibinfo {pages} {57} (\bibinfo {year} {1965})}\BibitemShut
  {NoStop}%
\bibitem [{\citenamefont {Hawking}(1975{\natexlab{b}})}]{hawking1975particle}%
  \BibitemOpen
  \bibfield  {author} {\bibinfo {author} {\bibfnamefont {S.~W.}\ \bibnamefont
  {Hawking}},\ }\href@noop {} {\bibfield  {journal} {\bibinfo  {journal}
  {Communications in mathematical physics}\ }\textbf {\bibinfo {volume} {43}},\
  \bibinfo {pages} {199} (\bibinfo {year} {1975}{\natexlab{b}})}\BibitemShut
  {NoStop}%
\bibitem [{\citenamefont {Israel}(1966)}]{Israel:1966rt}%
  \BibitemOpen
  \bibfield  {author} {\bibinfo {author} {\bibfnamefont {W.}~\bibnamefont
  {Israel}},\ }\href {https://doi.org/10.1007/BF02710419} {\bibfield  {journal}
  {\bibinfo  {journal} {Nuovo Cim. B}\ }\textbf {\bibinfo {volume} {44}},\
  \bibinfo {pages} {1} (\bibinfo {year} {1966})},\ \bibinfo {note} {[Erratum:
  Nuovo Cim.B 48, 463 (1967)]}\BibitemShut {NoStop}%
\bibitem [{\citenamefont {Hoegl}\ \emph {et~al.}(2020)\citenamefont {Hoegl},
  \citenamefont {Hofmann},\ and\ \citenamefont {Koegler}}]{Hoegl:2020hif}%
  \BibitemOpen
  \bibfield  {author} {\bibinfo {author} {\bibfnamefont {B.}~\bibnamefont
  {Hoegl}}, \bibinfo {author} {\bibfnamefont {S.}~\bibnamefont {Hofmann}},\
  and\ \bibinfo {author} {\bibfnamefont {M.}~\bibnamefont {Koegler}},\ }\href
  {https://doi.org/10.1103/PhysRevD.102.084065} {\bibfield  {journal} {\bibinfo
   {journal} {Phys. Rev. D}\ }\textbf {\bibinfo {volume} {102}},\ \bibinfo
  {pages} {084065} (\bibinfo {year} {2020})},\ \Eprint
  {https://arxiv.org/abs/2007.15717} {arXiv:2007.15717 [gr-qc]} \BibitemShut
  {NoStop}%
\bibitem [{\citenamefont {Mazharimousavi}\ \emph {et~al.}(2017)\citenamefont
  {Mazharimousavi}, \citenamefont {Halilsoy},\ and\ \citenamefont
  {Amen}}]{doi:10.1142/S0218271817501589}%
  \BibitemOpen
  \bibfield  {author} {\bibinfo {author} {\bibfnamefont {S.~H.}\ \bibnamefont
  {Mazharimousavi}}, \bibinfo {author} {\bibfnamefont {M.}~\bibnamefont
  {Halilsoy}},\ and\ \bibinfo {author} {\bibfnamefont {S.~N.~H.}\ \bibnamefont
  {Amen}},\ }\href {https://doi.org/10.1142/S0218271817501589} {\bibfield
  {journal} {\bibinfo  {journal} {International Journal of Modern Physics D}\
  }\textbf {\bibinfo {volume} {26}},\ \bibinfo {pages} {1750158} (\bibinfo
  {year} {2017})},\ \Eprint
  {https://arxiv.org/abs/https://doi.org/10.1142/S0218271817501589}
  {https://doi.org/10.1142/S0218271817501589} \BibitemShut {NoStop}%
\bibitem [{\citenamefont {Petrov}(1969)}]{Petrov}%
  \BibitemOpen
  \bibfield  {author} {\bibinfo {author} {\bibfnamefont {A.~Z.}\ \bibnamefont
  {Petrov}},\ }\href
  {https://www.elsevier.com/books/einstein-spaces/petrov/978-0-08-012315-8}
  {\emph {\bibinfo {title} {Einstein Spaces}}}\ (\bibinfo  {publisher}
  {Pergamon Press},\ \bibinfo {year} {1969})\BibitemShut {NoStop}%
\bibitem [{\citenamefont {Bunch}\ and\ \citenamefont
  {Parker}(1979)}]{bunchparker}%
  \BibitemOpen
  \bibfield  {author} {\bibinfo {author} {\bibfnamefont {T.~S.}\ \bibnamefont
  {Bunch}}\ and\ \bibinfo {author} {\bibfnamefont {L.}~\bibnamefont {Parker}},\
  }\href {https://doi.org/10.1103/PhysRevD.20.2499} {\bibfield  {journal}
  {\bibinfo  {journal} {Phys. Rev. D}\ }\textbf {\bibinfo {volume} {20}},\
  \bibinfo {pages} {2499} (\bibinfo {year} {1979})}\BibitemShut {NoStop}%
\end{thebibliography}%
	
\end{document}